\documentclass{emulateapj}
\usepackage{times}

\newcommand{\HST}{{\sl HST}}

\newcommand{\BI}{$B\!-\!I$}

\newcommand{\VI}{$V\!-\!I$}
\newcommand{\VpI}{$V\!+\!I$}

\newcommand{\ZH}{[$Z$/H]}

\newcommand{\picplace}[1]{\vbox{\hrule\@height 0.4pt\@width\hsize
\hbox to\hsize{\vrule\@width 0.4pt\@height#1\hfil
\vrule\@width 0.4pt\@height#1}\hrule\@height 0.4pt\@width\hsize}}

\slugcomment{Accepted by AJ}

\shorttitle{{\it HST/ACS\/} Imaging of Globular Clusters in NGC 3610}
\shortauthors{Goudfrooij et al.}

\begin{document}


\title{Dynamical Evolution of Globular Cluster Systems formed in
  Galaxy Mergers: \\ 
  Deep HST/ACS Imaging of Old and Intermediate-Age Globular
  Clusters in NGC~3610\altaffilmark{1}} 


\author{Paul Goudfrooij}
\affil{Space Telescope Science Institute, 3700 San Martin Drive,
  Baltimore, MD 21218; goudfroo@stsci.edu \vspace*{-3mm}} 


\author{Fran\c{c}ois Schweizer} 
\affil{Carnegie Observatories, 813 Santa Barbara Street, Pasadena, CA
  91101; schweizer@ociw.edu}

\and

\author{Diane Gilmore and Bradley C. Whitmore}
\affil{Space Telescope Science Institute, 3700 San Martin Drive,
  Baltimore, MD 21218; dkarakla@stsci.edu, whitmore@stsci.edu} 



\altaffiltext{1}{Based on observations with the NASA/ESA {\it Hubble
    Space Telescope}, obtained at the Space Telescope Science
    Institute, which is operated by AURA, Inc., under NASA contract
    NAS5--26555.} 


\begin{abstract}
The Advanced Camera for Surveys on board the {\it Hubble Space
    Telescope\/} has been used to obtain deep high-resolution images
    of the giant elliptical galaxy NGC~3610, a well-established remnant of
    a dissipative galaxy merger. These observations supersede
    previous, shallower observations which revealed the presence of a
    population of metal-rich globular clusters of intermediate age
    ($\sim$\,1.5\,--\,4 Gyr). We detect a total of 580 cluster candidates,
    46\% more than from the previous WFPC2 images.  
    The new photometry strengthens the significance of the previously found
    bimodality of the color distribution of clusters. Peak colors in \VI\ are
    0.93 $\pm$ 0.01  and 1.09 $\pm$ 0.01 for the blue and red subpopulations, 
    respectively. 
    The luminosity function of the inner 50\% of the
    metal-rich (`red') population of clusters differs
    markedly from that of the outer 50\%. In particular, the luminosity
    function of the inner 50\% of the red clusters shows a 
    flattening consistent with a turnover that is about 1.0 mag fainter than the
    turnover of the blue clusters. This is consistent with predictions of
    recent models of cluster disruption for the age range mentioned above and
    for metallicities that are consistent with the peak color of the red clusters
    as predicted by population synthesis models. The radial
    surface density profile of red clusters follows that of the
    underlying galaxy light more closely than in `normal' elliptical
    galaxies, which is consistent with the intermediate-age nature of the red
    clusters. We determine the specific frequency of clusters in NGC~3610 and
    find a present-day value of $S_N = 1.4 \pm 0.6$. Using published age
    estimates for the diffuse light of NGC~3610 as well as cluster disruption
    models, we estimate that this value will increase to $S_N = 3.8 \pm 1.7$
    at an age of 10 Gyr, which is consistent with typical $S_N$ values
    for `normal' elliptical galaxies. Our findings constitute further
    evidence in support of the notion that metal-rich cluster
    populations formed during major mergers involving gas-rich
    galaxies can evolve dynamically (through 
    disruption processes) into the red, metal-rich cluster populations that
    are ubiquitous in `normal' giant ellipticals.   
\end{abstract}




\section{Introduction}              \label{s:intro}
Recent deep imaging studies of `normal' giant elliptical galaxies with the
{\it Hubble Space Telescope (HST)\/} and large, ground-based telescopes
have shown that these galaxies usually contain rich globular cluster
(GC) systems with bimodal color distributions (e.g.,
\citealp*{zepash93,kunwhi01,lars+01,peng+06}). Typically, roughly half of 
the GCs
are blue, and half red. Follow-up spectroscopy with 8-m class telescopes
has revealed that both `blue' and `red' GC subpopulations are typically old
($\ga$\,8 Gyr, \citealp*{forb+01,cohe+03,puzi+05}) implying that the
bimodality is mainly due to differences in metallicity. The colors of the
`blue' GCs are usually similar to those of metal-poor halo GCs in the
Milky Way and M31, while the mean colors of the `red' GCs are
similar to those of the diffuse light of their host galaxies (e.g.,
\citealp*{geis+96,forb+97}). 
Hence, the nature of the `red', metal-rich GCs is likely to hold important
clues to the star formation history of their host galaxies. 

One environment {\it known\/} to produce metal-rich GCs and bimodal color
distributions is that of vigorous starbursts induced by mergers of gas-rich
galaxies. Massive young GCs have been commonly found in
mergers and young merger remnants using {\it HST\/} observations (e.g.,
\citealp*{holt+92,schw02}, and references therein). 
Follow-up spectroscopy has confirmed the ages (and in three cases even the high
masses, \citealp*{mara+04,bast+06}) of these young clusters predicted from their
colors and luminosities (e.g., \citealp{schsei98}). Their metallicities
tend to be near solar, as expected for clusters formed out of enriched gas
in spiral disks.  
A natural interpretation of these data is that the metal-rich GCs in 
`normal', old giant ellipticals formed in gas-rich mergers at $z \ga 1$,
and that the formation process of giant ellipticals with significant
populations of metal-rich GCs was similar to that in galaxy mergers
observed today \citetext{e.g., \citealp*{schw87,ashzep92}}. 

An often quoted hurdle for this `formation by
merger' scenario has been the marked difference in the luminosity functions
(hereafter LFs) of old vs.\ young GC systems (e.g., \citealp{vdb95a}).  
The LFs of old GC systems of `normal' galaxies are well fit by 
Gaussians peaking at $M_V^0 \simeq -7.2$ mag with a dispersion of
$\sigma \simeq 1.3$ mag \citetext{e.g., \citealp{whit97}}, while young GC
systems in mergers and young remnants have power-law LFs with indices of
$\alpha \simeq -2$ \citetext{e.g., \citealp{whit+99}}. 
Intermediate-age merger remnants with ages of 1\,--\,5 Gyr
are ideal probes for studying the long-term dynamical effects on GC systems
formed during a major merger. Such galaxies are still 
identifiable as merger remnants through their morphological fine structure 
\citep[e.g.,][]{schsei92}, yet they are old enough to ensure that substantial
dynamical evolution of the globular clusters has already occurred. 
Recent calculations of 
dynamical evolution of GCs (including two-body relaxation, tidal shocking,
and stellar mass loss) show that the least massive GCs are expected to
disrupt first as galaxies age, which can gradually transform the
initial power-law LFs into LFs with Gaussian-like peaks or turnovers 
\citep[e.g.,][]{vesp98,baum98,falzha01,veszep03,prigne06}.

Recent \HST\ studies of candidate intermediate-age merger remnants have
revealed that their `red' GC subpopulations show LFs consistent with power 
laws (as expected if formed during a recent merger event;
\citealp*{goud+01b,whit+02}, hereafter W02). However, the
combination of field of view and sensitivity of the \HST\ instrumentation was not
quite efficient enough to allow a detection of a turnover
in the LFs until just a few years ago. The first empirical evidence for 
dynamical evolution of red GC LFs from a power law toward a Gaussian-like shape
in an intermediate-age early-type galaxy was facilitated by the unprecedented
sensitivity of the Advanced Camera for Surveys {\it (ACS)\/} aboard {\it
  HST}. Based on {\it ACS\/} observations, \citet{goud+04} found that
the outer half of  the red GC system of the 3-Gyr-old merger remnant
NGC~1316 still shows a power-law LF, whereas the inner half shows a
LF with a turnover characteristic of old GC systems.  

In an effort to test how common this result may be among nearby
intermediate-age ellipticals, the current paper reports on new {\it
  ACS\/} imaging of NGC~3610.  
Several optical imaging studies have shown that NGC~3610 is a bona fide 
inter\-medi\-ate-age merger remnant. Its colors are anomalously blue for a 
giant elliptical galaxy of its luminosity
\citep*{schsei92,goud+94a,idia+02}, suggesting recent star formation  
activity. Its inner region features a small, strongly twisted inner disk
\citep{scoben90,whit+97}. Such disks are a natural byproduct of gas-rich galaxy
mergers in simulations \citep{barn02}. The galaxy's outer envelope is extremely boxy
relative to its inner body and shows several non-concentric plumes and tails
\citep{seisch90} which are most likely remnants of tidal perturbations. In
fact, NGC~3610 has the highest ``fine structure index'' (indicative of
structures believed to be resulting from galactic mergers) of all 74
early-type galaxies in the sample of \citet{schsei92}. Finally, its spheroid 
is well fit by a de Vaucouleurs' $r^{1/4}$ profile and has a
surprisingly small effective radius for the galaxy's luminosity \citep[13$''$ or 2.1 kpc
in the $B$ band,][]{burs+87}. 
Overall, these features are consistent with NGC~3610 being the product of a
dissipative merger with incomplete dynamical relaxation. 

In this paper we adopt a distance of 33.9 Mpc for NGC~3610,
corresponding to a distance modulus of $(m-M)_0 = 32.65$ mag as
measured from surface brightness fluctuations \citep{tonr+01}. 
The Galactic foreground reddening towards NGC~3610 is $A_V$ = 0.00 and
0.03 according to \citet{burhei84} and \citet{schl+98},
respectively. In accordance with most literature on photometry of NGC~3610
\citep[e.g.,][]{goud+94a,whit+97,whit+02,mich05}, we adopt the
\citet{burhei84} value in this paper. We do not make any corrections
for reddening internal to NGC~3610, given the absence of any patchy
absorption or color changes in optical images of NGC~3610
\citep{goud+94b,whit+97} and the non-detection of far-infrared
emission \citep{knap+89}.

\section{Observations} \label{s:obs}

NGC~3610 was observed with {\it HST\/} on June 29th, 2003, using the
wide-field channel (WFC) of {\it ACS\/} and the F435W, F555W, and F814W filters, with
total exposure times of 1080 s, 6410 s, and 6060 s, respectively. 
The F555W and F814W exposure times were chosen to reach a similar depth for
an object of \VI\ = 1.0, the mean \VI\ color of globular cluster systems of
giant elliptical galaxies \citep[e.g.,][]{kunwhi01,lars+01}. 
The F435W exposures were mainly included to provide a large color
baseline to allow accurate age/metallicity measurements for the bright GCs, 
which does not require reaching very faint levels. Hence, the F435W exposures were
relatively short and taken at only two positions (``dithers'') offset from
each other by 3.5 columns of the WFC CCD. 
In contrast, the final F555W and F814W images were constructed from long (19\,--\,26
min) exposures taken at 5 dither positions, supplemented by one 
short exposure taken in each filter to avoid saturation of the bright
central regions.  
The individual images in each band were combined using the task {\sc
  multidrizzle} within {\sc PyRAF/stsdas v3.3}, with task parameters
set to {\it pixfrac\/} = 0.8 and {\it scale\/} = 1.0. 
Saturated pixels in the long exposures were replaced by those in the short
exposures while running {\sc multidrizzle} by setting the appropriate data
quality flag in the affected pixels of the long exposures.  
The gain in sensitivity, spatial resolution, and spatial coverage of
the new {\it ACS\/} images relative to the available {\it WFPC2\/}
images allowed us to detect roughly 50\% more GC candidates than
the {\it WFPC2\/} images used in W02 did. 

\section{Cluster Selection and Photometry}          \label{s:selphot}

\subsection{Cluster Selection}    \label{s:sel}

Prior to performing source selection and photometry, the strongly
varying light distribution of the galaxy was fitted and removed in order
to minimize errors in the photometry due to any particular choice of
object aperture and sky annulus (see below). Experimentation
with various fitting methods revealed that {\it any\/} model fit to the galaxy
surface brightness distribution is strongly hampered by the presence
of a very  prominent stellar disk in the central regions
\citep[e.g.,][]{scoben90,goud+94a,whit+97,whit+02}. Best results were
achieved by approximating the diffuse light of the galaxy by applying a 
median filter with a 41\,$\times$\,41 pixel$^2$ kernel to the image. 

The first step in the selection of candidate GC targets was the
application of the {\sc daofind} task from the {\sc daophot-ii\/} package
\citep{stet87} to an image prepared by dividing the sum of the F555W
and F814W images (i.e., the \VpI\ image) by the square root of the
corresponding median-filtered image (thus yielding uniform shot noise
characteristics over the whole image). The detection threshold was set
at 3\,$\sigma$ above the background. The region around the central
stellar disk was masked out in this step. To allow the detection of
candidate GCs in the region occupied by the stellar disk, {\sc
  daofind} was applied again on the \VpI\ image divided by the square
root of a median-filtered version of it, but this time using a much smaller
kernel (7\,$\times$\,7 pixel$^2$).  From this last run of {\sc
  daofind}, we only included objects detected in the central area that was
masked out in the first run into the final object list.   

At this stage, a total of 1472 objects were detected. 
Task {\sc phot} was then applied to obtain fixed-aperture photometry for
all those objects in the median-subtracted $B$, $V$, $I$, and \VpI\ images,
employing several aperture radii. Due to the substantial structure in the
median-filtered images in the innermost region, aperture photometry for the objects in
the innermost region was obtained from the direct images, employing small
apertures (radii for object/inner background/outer background were set at
2/3/5 pixels in this case; the photometry of these inner objects is shown
with open symbols in Figure~\ref{f:cmd_hists}). 

\begin{figure}
\plotone{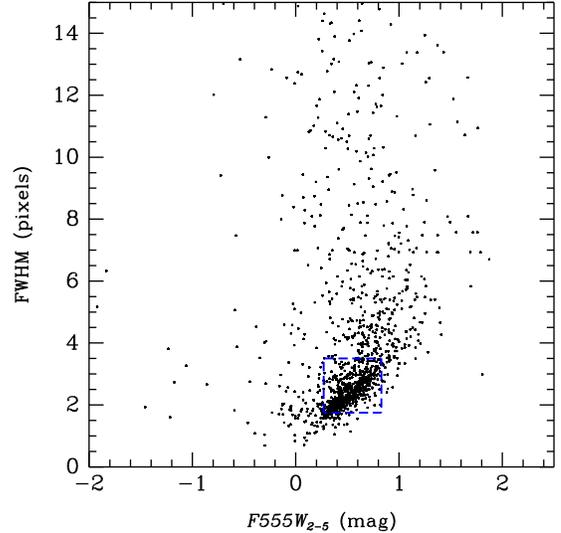}
\caption{Illustration of the size selection procedure. The FWHM of the
  object profile measured as discussed in Sect.~\ref{s:sel} is plotted
  versus F555W$_{2-5}$, the difference in F555W magnitudes measured using
  aperture radii of 2 and 5 pixels. The region within dashed lines
  represents the parameter space used to select GC candidates in NGC~3610.  
  \label{f:sizeplot}}  
\end{figure}

Since the object list at this point still includes Galactic foreground stars,
background galaxies, and any residual cosmic rays and/or hot pixels in
addition to the star clusters we are interested in studying, we 
devised an automated selection procedure based on two size-related parameters
and the \VI\ color. 
Quantitative size selection criteria were developed based on globular
clusters known to be associated with NGC~3610 by means of radial 
velocities \citep*{stra+04}, obvious bright foreground stars with
visible diffraction spikes, and visual inspection of objects (especially
near the extrema of the selection criteria). 
The first size parameter used is a `concentration index', defined
as the difference in F555W magnitudes measured using aperture radii of 2 and 5
pixels, hereafter denoted as F555W$_{2-5}$. 
Visual inspection of the objects remaining after restricting F555W$_{2-5}$
to the range 0.29 -- 0.83 mag showed that many extended
background galaxies, all obvious hot pixels, and some bright foreground
stars were filtered out. However, this selection still left a
substantial number of background galaxies in the target list. 
The second size parameter used for the classification is the FWHM of the
radial profile of the object. Rather than assuming a specific model of the
radial profile (e.g., Gaussian or Moffat), we used a more robust, direct
measure of the FWHM by computing the radial derivative of the azimuthally
averaged enclosed flux profile $F(R)$: 
\[ F(R) \, = \, \int_0^R P(r) \,r\,\mbox{d}r \]
where $P(r)$ is the radial profile of intensity per unit area. The peak of
$d F(R)/R\,dr$ is found and the FWHM is twice the radius of the profile at
half the peak value.   
As illustrated in Figure~\ref{f:sizeplot}, we selected objects with
$0.29 \leq {\rm F555W}_{2-5} \leq 0.83$ mag and $1.75  \leq \mbox{FWHM (pixels)} \leq
  3.50$ in this step.  

At the distance of NGC~3610, typical GCs are only marginally resolved at
the resolution of the drizzled ACS images. This makes it somewhat difficult to
distinguish foreground stars from GC candidates based on size parameters
alone, especially at faint magnitudes where the absence of diffraction
spikes hampers the visual identification of objects as foreground
stars. Hence we use the \VI\ color as another discriminant between stars
and GC candidates by applying a color selection cut of $0.5 < (V\!-\!I)_0
< 1.5$. This interval constitutes the full range in \VI\ expected for GCs
older than 1 Gyr and $-2.5 < [Z/{\rm H}] < 0.5$ according to the simple
stellar population (SSP) models of \citet{bc03} and \citet{mara05}, plus 
0.2 mag of padding on the blue and red ends to allow for reasonable
photometric errors.  These concentration, size, and color cuts yielded a list of 
580 GC candidates. Any further contamination of this list by foreground stars 
and compact background galaxies will be addressed by statistical subtraction 
(see below). 

\subsection{Photometry}  \label{s:phot}

Aperture corrections were $-0.26 \,(\pm0.01)$ mag in F435W $(B)$,
$-0.24\,(\pm0.01)$ in F555W $(V)$ and $-0.29 \,(\pm0.01)$
mag in F814W $(I)$ for an aperture radius of 3 WFC pixels (i.e.,
0\farcs15). These values were determined through measurements of several
isolated, well-detected GC candidates located throughout the field. These GC
candidates spanned the range of size parameters mentioned in the previous
section. We considered correcting for the effect of imperfect charge transfer
efficiency (CTE). However, the effect was found to be negligible ($<$\,0.01
mag) according to the formulae of \citet{riemac04}, even at the corners of
the F435W image which have the lowest background level by far ($\sim$\,30
e$^-$/pixel). A correction for CTE loss was therefore not applied. In
order to facilitate comparisons of our photometry with predictions of
various SSP models, we transformed the instrumental magnitudes to
Johnson/Cousins $B$, $V$ and $I$ using the calibrations derived by the
ACS Instrument Definition Team \citep{siri+05}. Photometry and
astrometry for the 50 brightest GC candidates in the ACS frames are
listed in Table~\ref{t:phot}. 


\begin{table*}
\begin{center}
\footnotesize
\caption{Photometry and astrometry of the 50 brightest GC candidates in NGC~3610 
 from the ACS data. The object list is sorted on $V$ magnitude (brightest first). 
 \label{t:phot}}
\begin{tabular*}{14cm}{@{\extracolsep{\fill}}lrcccrrr@{}}
\multicolumn{3}{c}{~} \\ [-2.5ex]   
 \tableline \tableline
\multicolumn{3}{c}{~} \\ [-2.2ex]                                                
Object & W02 & $V$ & \VI\ & \BI\ & \multicolumn{1}{c}{$\Delta$ RA} & \multicolumn{1}{c}{$\Delta$ DEC} & 
 Radius \\ [0.2ex]
ID & ID & (mag) & (mag) & (mag) & \multicolumn{1}{c}{(arcsec)} & \multicolumn{1}{c}{(arcsec)} & 
 (arcsec)  \\ [0.5ex] \tableline 
\multicolumn{3}{c}{~} \\ [-2.ex]              
  G1 & 2  & 21.530 $\pm$ 0.498 &   1.095 $\pm$   0.784 & 1.971 $\pm$ 0.804 &      0.31 &      1.23 &    1.30 \\
  G2 & 3  & 21.548 $\pm$ 0.007 &   1.062 $\pm$   0.010 & 1.894 $\pm$ 0.015 &     21.35 &      5.07 &   22.21 \\
  G3 & ---& 21.567 $\pm$ 0.007 &   1.047 $\pm$   0.010 & 1.822 $\pm$ 0.015 &  $-$39.73 &     38.77 &   56.18 \\
  G4 & 6  & 21.659 $\pm$ 0.007 &   1.054 $\pm$   0.010 & 1.835 $\pm$ 0.015 &     21.86 &   $-$5.88 &   22.90 \\
  G5 & 5  & 21.668 $\pm$ 0.054 &   1.068 $\pm$   0.056 & 1.867 $\pm$ 0.057 &   $-$4.93 &      2.14 &    5.45 \\
  G6 & 16 & 21.872 $\pm$ 0.323 &   1.191 $\pm$   0.561 & 1.910 $\pm$ 0.536 &   $-$0.20 &      1.92 &    1.96 \\
  G7 & 14 & 21.887 $\pm$ 0.312 &   1.040 $\pm$   0.561 & 2.169 $\pm$ 0.708 &      1.90 &   $-$0.00 &    1.91 \\
  G8 & ---& 21.894 $\pm$ 0.008 &   0.931 $\pm$   0.011 & 1.270 $\pm$ 0.015 &  $-$59.39 &   $-$0.13 &   60.11 \\
  G9 & 9  & 21.955 $\pm$ 0.008 &   1.066 $\pm$   0.011 & 1.876 $\pm$ 0.018 &     67.05 &  $-$17.14 &   70.04 \\
 G10 & 11 & 21.983 $\pm$ 0.008 &   1.125 $\pm$   0.012 & 2.007 $\pm$ 0.021 &  $-$10.97 &   $-$8.23 &   13.89 \\
 G11 & 12 & 22.011 $\pm$ 0.008 &   0.950 $\pm$   0.012 & 1.654 $\pm$ 0.018 &   $-$5.93 &     15.78 &   17.06 \\
 G12 & 15 & 22.038 $\pm$ 0.423 &   1.076 $\pm$   0.757 & 1.920 $\pm$ 0.765 &      1.62 &      0.37 &    1.68 \\
 G13 & 10 & 22.046 $\pm$ 0.008 &   0.893 $\pm$   0.012 & 1.522 $\pm$ 0.018 &     60.15 &  $-$55.51 &   82.83 \\
 G14 & 17 & 22.076 $\pm$ 0.009 &   1.031 $\pm$   0.012 & 1.808 $\pm$ 0.019 &     16.57 &  $-$15.51 &   22.97 \\
 G15 & 7  & 22.082 $\pm$ 0.009 &   1.105 $\pm$   0.012 & 1.920 $\pm$ 0.021 &     17.42 &   $-$3.81 &   18.04 \\
 G16 & 13 & 22.115 $\pm$ 0.011 &   1.122 $\pm$   0.014 & 1.936 $\pm$ 0.023 &      3.84 &   $-$5.55 &    6.83 \\
 G17 & 18 & 22.433 $\pm$ 0.011 &   1.143 $\pm$   0.014 & 1.980 $\pm$ 0.025 &      7.40 &      3.21 &    8.15 \\
 G18 & 20 & 22.463 $\pm$ 0.010 &   0.934 $\pm$   0.014 & 1.607 $\pm$ 0.022 &  $-$32.92 &  $-$37.30 &   50.35 \\
 G19 & 19 & 22.569 $\pm$ 0.061 &   1.239 $\pm$   0.056 & 2.079 $\pm$ 0.061 &      5.29 &      0.84 &    5.41 \\
 G20 & 22 & 22.597 $\pm$ 0.011 &   0.896 $\pm$   0.015 & 1.583 $\pm$ 0.024 &     54.59 &  $-$55.51 &   78.79 \\
 G21 & 24 & 22.598 $\pm$ 0.244 &   1.131 $\pm$   0.320 & 2.852 $\pm$ 0.691 &   $-$0.43 &      3.01 &    3.08 \\
 G22 & 25 & 22.615 $\pm$ 0.011 &   0.913 $\pm$   0.015 & 1.569 $\pm$ 0.024 &  $-$47.25 &  $-$50.24 &   69.79 \\
 G23 & 26 & 22.662 $\pm$ 0.013 &   1.059 $\pm$   0.017 & 1.913 $\pm$ 0.029 &   $-$9.11 &      3.69 &    9.96 \\
 G24 & 23 & 22.685 $\pm$ 0.013 &   1.091 $\pm$   0.017 & 1.872 $\pm$ 0.027 &   $-$0.23 &      8.39 &    8.49 \\
 G25 & 27 & 22.801 $\pm$ 0.380 &   1.196 $\pm$   0.509 & 1.982 $\pm$ 0.479 &      3.38 &   $-$2.71 &    4.37 \\
 G26 & 28 & 22.836 $\pm$ 0.013 &   0.984 $\pm$   0.017 & 1.783 $\pm$ 0.028 &   $-$3.82 &  $-$77.72 &   78.76 \\
 G27 & ---& 22.852 $\pm$ 0.013 &   1.133 $\pm$   0.017 & 2.011 $\pm$ 0.029 &  $-$56.78 &  $-$32.04 &   65.98 \\
 G28 & ---& 22.882 $\pm$ 0.013 &   0.910 $\pm$   0.017 & 1.555 $\pm$ 0.027 &  $-$51.21 &  $-$62.83 &   82.03 \\
 G29 & ---& 22.938 $\pm$ 0.013 &   1.047 $\pm$   0.017 & 1.806 $\pm$ 0.029 &  $-$36.19 &   $-$6.62 &   37.24 \\
 G30 & 31 & 23.070 $\pm$ 0.014 &   1.059 $\pm$   0.019 & 1.879 $\pm$ 0.032 &      9.71 &  $-$41.71 &   43.35 \\
 G31 & 32 & 23.094 $\pm$ 0.014 &   1.053 $\pm$   0.019 & 1.801 $\pm$ 0.031 &     58.58 &  $-$67.12 &   90.15 \\
 G32 & 30 & 23.120 $\pm$ 0.015 &   1.018 $\pm$   0.021 & 1.759 $\pm$ 0.033 &     37.40 &     17.47 &   41.76 \\
 G33 & ---& 23.166 $\pm$ 0.015 &   1.111 $\pm$   0.020 & 1.978 $\pm$ 0.035 &     14.36 &     96.98 &   99.23 \\
 G34 & 33 & 23.175 $\pm$ 0.015 &   1.072 $\pm$   0.019 & 1.853 $\pm$ 0.033 &      3.75 &  $-$28.17 &   28.77 \\
 G35 & 34 & 23.197 $\pm$ 0.017 &   1.071 $\pm$   0.021 & 1.863 $\pm$ 0.036 &  $-$10.59 &     12.36 &   16.48 \\
 G36 & 35 & 23.340 $\pm$ 0.017 &   1.121 $\pm$   0.022 & 1.923 $\pm$ 0.037 &      2.44 &  $-$18.53 &   18.91 \\
 G37 & 37 & 23.347 $\pm$ 0.021 &   1.101 $\pm$   0.025 & 1.951 $\pm$ 0.044 &      0.02 &   $-$8.52 &    8.61 \\
 G38 & 38 & 23.384 $\pm$ 0.020 &   0.977 $\pm$   0.027 & 1.650 $\pm$ 0.042 &       0.54 &  $-$8.78 &    8.90 \\
 G39 & 39 & 23.391 $\pm$ 0.333 &   1.372 $\pm$   0.366 & 2.168 $\pm$ 0.371 &   $-$2.57 &      4.28 &    5.06 \\
 G40 & 36 & 23.403 $\pm$ 0.018 &   1.153 $\pm$   0.024 & 2.057 $\pm$ 0.043 &   $-$9.68 &  $-$18.34 &   20.99 \\
 G41 & ---& 23.429 $\pm$ 0.017 &   0.954 $\pm$   0.023 & 1.688 $\pm$ 0.037 &  $-$53.47 &     66.51 &   86.38 \\
 G42 & ---& 23.516 $\pm$ 0.018 &   0.890 $\pm$   0.023 & 1.547 $\pm$ 0.037 &  $-$39.71 &     23.19 &   46.55 \\
 G43 & 41 & 23.550 $\pm$ 0.018 &   0.925 $\pm$   0.023 & 1.591 $\pm$ 0.038 &      8.43 &  $-$72.84 &   74.21 \\
 G44 & 43 & 23.560 $\pm$ 0.019 &   1.166 $\pm$   0.024 & 2.009 $\pm$ 0.043 &  $-$12.33 &  $-$10.03 &   16.08 \\
 G45 & ---& 23.597 $\pm$ 0.019 &   1.114 $\pm$   0.024 & 1.936 $\pm$ 0.041 &  $-$27.75 &     43.83 &   52.51 \\
 G46 & 40 & 23.598 $\pm$ 0.019 &   1.106 $\pm$   0.025 & 1.913 $\pm$ 0.043 &  $-$13.11 &  $-$59.10 &   61.26 \\
 G47 & 49 & 23.599 $\pm$ 0.019 &   1.119 $\pm$   0.023 & 1.918 $\pm$ 0.039 &     48.15 &  $-$29.76 &   57.26 \\
 G48 & 42 & 23.636 $\pm$ 0.019 &   1.057 $\pm$   0.026 & 1.909 $\pm$ 0.047 &      3.30 &  $-$31.04 &   31.58 \\
 G49 & 45 & 23.640 $\pm$ 0.020 &   1.064 $\pm$   0.026 & 1.927 $\pm$ 0.048 &      4.92 &  $-$20.53 &   21.37 \\
 G50 & 44 & 23.697 $\pm$ 0.020 &   1.126 $\pm$   0.026 & 1.989 $\pm$ 0.046 &     30.48 &   $-$1.42 &   30.89 \\ [0.5ex]
 \tableline
\end{tabular*}
\tablecomments{Columns $\Delta$~RA and $\Delta$~DEC give the 
positional offsets of the GC candidates from the center of NGC~3610 in RA and 
DEC, respectively. The center of NGC~3610 in the ACS data is (RA, DEC) = 
(11$^{\rm d}$18$^{\rm m}$25\fs375, +58$^{\circ}$47$'$10\farcs75) in J2000.0 
equinox.}
\end{center}
\end{table*}

Considering the sky coverage common to this study and that of W02, a
comparison of the photometry in Table~\ref{t:phot} with that of the brightest
50 GC candidates in the WFPC2 data of W02 (i.e., their Table 1) shows a
weighted mean difference (in the sense ``our minus W02'') in $V$ magnitudes of
$-$0.01 with an rms scatter of 0.12 mag. (The weighting was done by means of the
inverse variance.) The difference in weighted mean \VI\ colors is $-$0.02
with a rms scatter of 0.11 mag. This comparison excludes objects \# 1, 4, and
8 of W02, which are all located within the central disk of NGC~3610. A
careful check of the ACS images does not show any evidence for the
presence of objects on top of the bright disk light at these
locations.   

\begin{figure}
\plotone{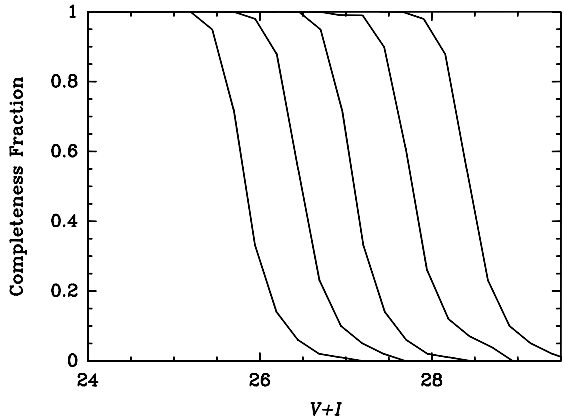}
\caption{Completeness functions used for the ACS photometry of
NGC~3610 GC candidates. Lines are shown for five background levels: From
left to right: 45000, 14250, 4500, 1425, and 450 e$^{-}$ 
pixel$^{-1}$ for an \VpI\ exposure time of 2360 s.\label{f:completeness}}  
\end{figure}

Completeness corrections were performed using the \VpI\ image, i.e., the
one used to determine the initial list of GC candidates, using routines in the
{\sc daophot} package. 
We added artificial objects (in batches of 100) with
a radial intensity profile derived from a fit to real GC candidates in the
frame to the \VpI\ image, for five different background levels and several
magnitude intervals. The color of the objects was set equal to the
median color of GC candidates. Prior to running {\sc daofind} to determine the
fraction of retrieved artificial objects, the image was divided by the
square root of its median-filtered version, as was done during the initial
object finding procedure. Artificial objects
retrieved by {\sc daofind} were also subjected to
the size selection criteria mentioned above using the
median-filter-subtracted \VpI\ image before including them in the
object counts\footnote{To reflect the different method of
object photometry in the innermost regions (see Section~\ref{s:sel}),
the photometry of the artificial objects at the highest background
level was done on the direct image instead of the
median-filter-subtracted image.}. 
Figure \ref{f:completeness} shows the resulting completeness functions. 

\begin{figure}
\plotone{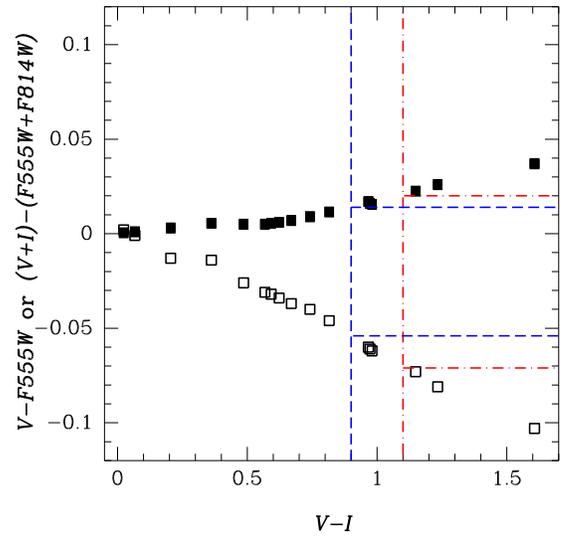}
\caption{The relation of the photometric zeropoints of $V$ (open squares)
  and \VpI\ (filled squares) with the \VI\ color. Note that the dependence
  of the \VpI\ zeropoint on \VI\ color is significantly smaller than that of the
  $V$ zeropoint. Dashed lines show the zeropoints for typical `blue'
  globular clusters found in NGC~3610 (\VI\ $\sim$ 0.9), and dash-dot lines
  do so for typical `red' globular clusters in NGC~3610 (\VI\ $\sim$
  1.1). See discussion in Sect.~\ref{s:phot}. \label{f:vplusiplot}}   
\end{figure}

The two main reasons for using the \VpI\ image to determine the initial
source list and completeness corrections (rather than the $V$ or $I$ image,
which is often done) are {\it (i)\/} the greater depth reached by the 
\VpI\ image, and {\it (ii)\/} the fact that the 
transformation from instrumental magnitudes to Johnson/Cousins 
\VpI\ is significantly less dependent on source color than the
transformations to either $V$ or $I$, 
which helps during the completeness correction. The latter
point is illustrated in Figure~\ref{f:vplusiplot} which shows the dependence
of the photometric zeropoints for $V$ and \VpI\ on the \VI\ color. The zeropoints were
derived using the {\sc synphot} package within {\sc iraf/stsdas}. Synthetic
spectra of stellar types ranging from F2V to M0III from the Bruzual \& Charlot
(\citeyear{bc93}) library were convolved with Johnson/Cousins $V$ and $I$
filter passbands to yield Vega-based magnitudes, and with the ACS/WFC F555W
and F814W passbands to yield STMAG magnitudes. As Figure~\ref{f:vplusiplot}
shows, the dependence of the \VpI\ zeropoint on \VI\ color is much smaller 
than that of the $V$ zeropoint. In the context of the two common subpopulations of
clusters in giant ellipticals, which typically have \VI\ $\sim$ 0.95 and
1.2 for the blue and red subpopulations, respectively, the difference in
\VpI\ zeropoint is only 0.006 mag (vs.\ 0.016 mag for the $V$ zeropoint). Hence, the 
determination of completeness fractions using \VpI\ needs only be conducted for one
object color\footnote{Note however that these advantages for using the summed
  $(V\!+\!I)$ image are only applicable if the individual $V$ and $I$
images are of similar depth for the median color of the objects of
interest.}. 

\section{Properties of the Globular Cluster System} \label{s:gcprops}

\subsection{Color Distributions} \label{s:colors}

\begin{figure*}
\epsscale{0.9}
\plottwo{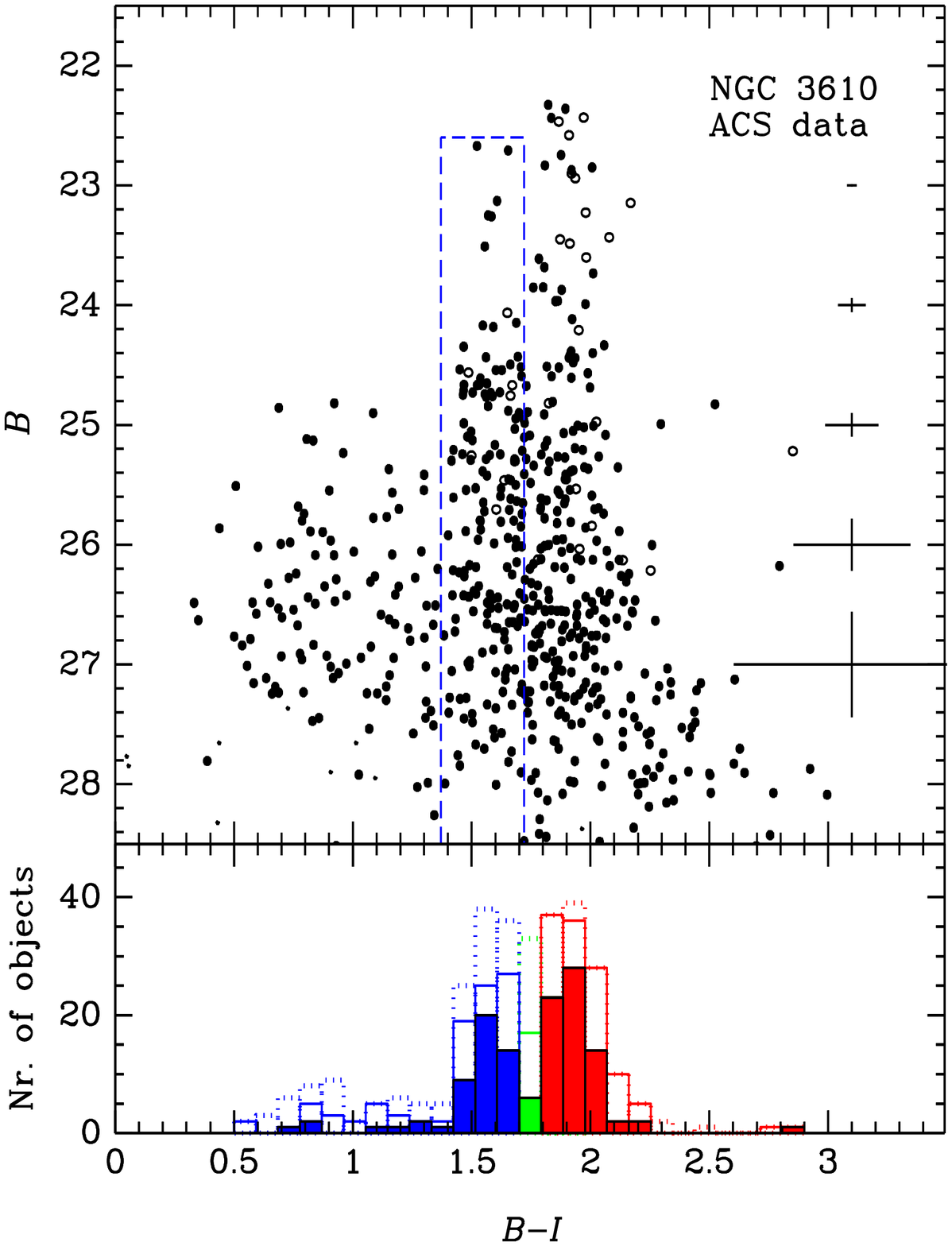}{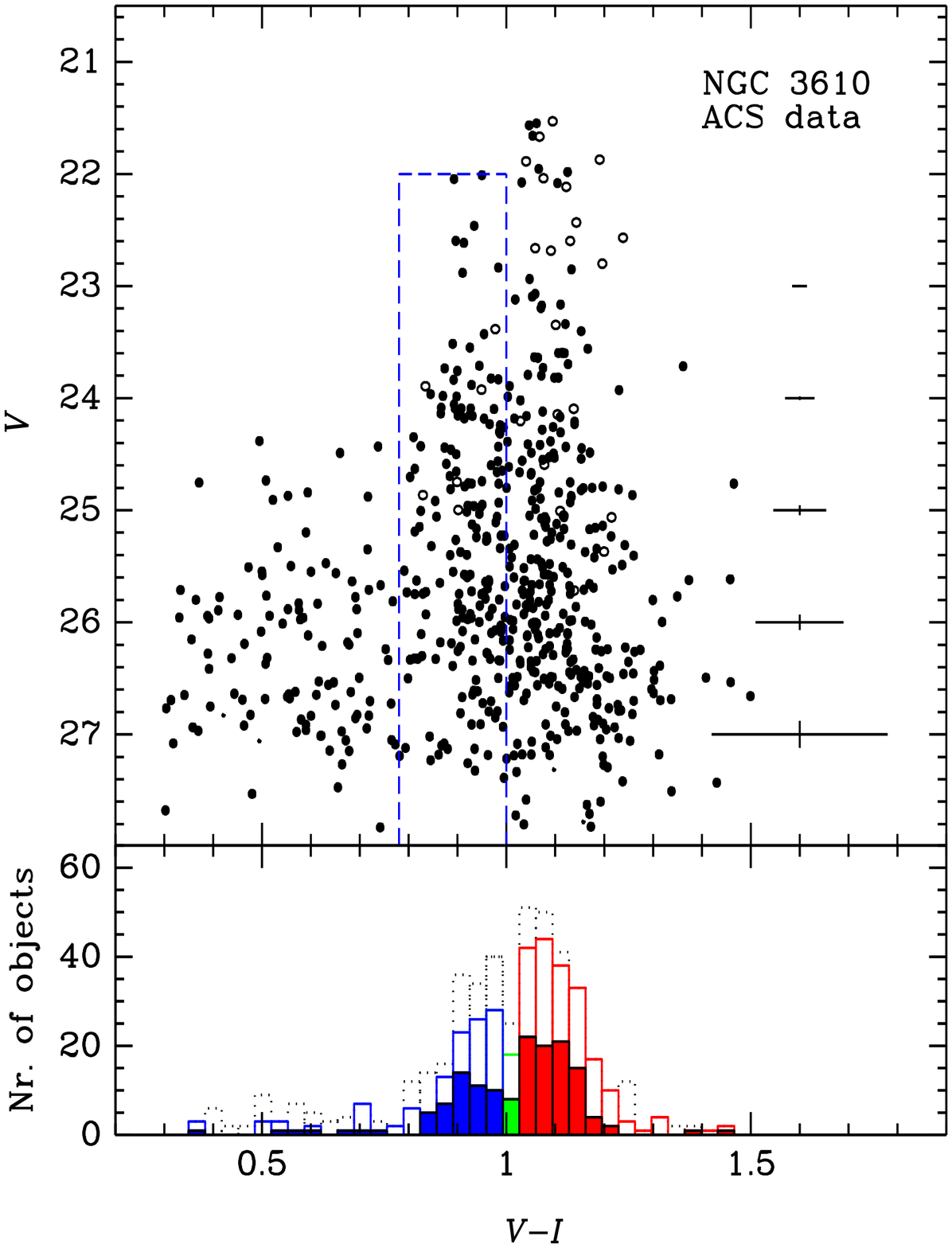}
\caption{
{\it Top panels}: $B$ vs.\ \BI\ (left) and $V$ vs.\ \VI\ (right)
  color-magnitude diagrams for GC candidates in
  NGC~3610. 
The GC candidates with projected galactocentric radii $< 12''$ are shown 
as open circles, while the others are shown as filled circles.  
Regions within dashed lines represent the magnitude and color
ranges for old halo GCs in the Milky Way, placed at the distance of NGC~3610. 
Representative error bars are shown on the right-hand side of the diagrams.  
{\it Bottom panels}: Distributions of \BI\ (left) and \VI\ (right)
  colors for GC candidates in NGC~3610. The bottom left panel shows 
  all GCs with $B \leq 26.5$ (open histogram), and GCs with 
  $B \leq 25.5$ (filled histogram). The dashed lines represent GCs 
  with $B \leq 26.5$ before correction for contamination by compact
  background galaxies. The corresponding histograms in the bottom right
  panel are for $V \leq 26.5$ (open histogram and dashed lines) and for $V
  \leq 25.0$ (filled histogram).\label{f:cmd_hists}}    
\end{figure*}

The top panels of Figure~\ref{f:cmd_hists} show the $B$ vs.\ \BI\
and $V$ vs.\ \VI\ color-magnitude diagrams (hereafter CMDs) for GC
candidates in NGC~3610. For reference, the region occupied by old halo GCs
in our Galaxy \citep[taken from the database of][]{harr96}, when placed
at the distance of NGC~3610, is indicated by dashed lines in both CMDs.   

The CMDs show a color distribution of GC candidates that is clearly
bimodal down to $B \sim 26.5$ and $V \sim 26.5$ where the typical color 
uncertainties are $\pm 0.22$ mag in \BI\ and $\pm 0.10$ mag in \VI. The
`blue' sub-population of GCs has colors and luminosities that are  
consistent with those of old halo GCs in our Galaxy, while the (larger)
`red' population has a mean color that is $\sim$\,0.4 mag redder than
that of the `blue' sub-population. The mean colors of each subpopulation seem
constant throughout the sampled range of luminosities, taking the color
uncertainties as a function of brightness into account. 
There is no obvious evidence for the `red slant' effect seen among `blue' GCs
in brightest cluster ellipticals, i.e., the effect that the mean color of
`blue' GCs gets progressively redder (indicating a higher metallicity) with
increasing GC luminosity \citep{harr+06,stra+06}.  
The $V$ vs.\ \VI\ CMD looks quite similar to the corresponding diagram in W02
(i.e., their Figure 3), except that the color bimodality seems more
obvious in the current CMD.  

The \BI\ and \VI\ color distributions of the GC candidates are shown in the
bottom panels of Figure~\ref{f:cmd_hists}, for two brightness cuts. Prior to
plotting, the color histograms of candidate GCs were corrected for
contamination by compact background objects. This correction was derived
from the compact objects detected beyond a projected radius of 100$''$ from
the galaxy center. The color histogram of the outer compact objects was
smoothed with a five-bin kernel to diminish small-number noise and scaled to
the area of the full ACS image prior to subtraction from the color histogram
of the full ACS image. The effect of this correction is shown in
Figure~\ref{f:cmd_hists}. While the surface number density of GC candidates
beyond 100$''$ from the galaxy center seems to be close to an asymptotic
background value (see Sect.~\ref{s:numdens} below), there may still be a
(small) number of true halo GCs included in the number density estimate of
compact background galaxies. Hence the correction for compact background
galaxies may be slightly overestimated, and the true color distribution of
GCs in NGC~3610 may lie slightly above (but most likely very close to)
the solid lines in the bottom panels of Figure~\ref{f:cmd_hists}.    

The unbiased probability of a bimodal color distribution (as opposed to a
unimodal one) was estimated using the Kaye's Mixture Model (KMM) test
\citep*{mclbas88,ashm+94}, by comparing fits of one Gaussian vs.\ two
Gaussians to the color distribution using the ``expectation-maximization''
(EM) method. To gain leverage in the presence of noisy data, we consider the
homoscedastic case of the KMM test where the standard deviation $\sigma$ is
the same for both Gaussians. Using the KMM test on various artificial color
distributions, we determine that a two-Gaussian fit is a better fit to the
data than a one-Gaussian fit if the $p$-value of the two-Gaussian fit is less
than 0.05. We ran the KMM test on several subsets of the color distributions
of GC candidates, using two magnitude limits and two upper limits of
the projected galactocentric radius. 
Table~\ref{t:kmmresults} lists the results of the KMM
tests. In all cases the distributions are judged to be bimodal at the
96\,--\,100\% probability level. The mean colors of the blue peak are \BI\ =
$1.61\pm 0.01$ and \VI\ = $0.93 \pm 0.01$, and those of the red peak are \BI\
= $1.92 \pm 0.01$ and \VI\ = $1.09 \pm 0.01$.  

\begin{table*}
\begin{center}
\caption{Results of the Kaye's Mixture Model (KMM) tests to the color
  distributions \label{t:kmmresults}}  
\small
\begin{tabular}{@{}crrccccccc@{}}
\multicolumn{3}{c}{~} \\ [-2.5ex]   
\tableline \tableline
\multicolumn{3}{c}{~} \\ [-1.8ex]                                                
Magnitude & $R_{\rm out}$\tablenotemark{a} & N$_{\rm obs}$\tablenotemark{b} & 
 $\sigma$\tablenotemark{c} & $(B\!-\!I)_{\rm blue}$ & $(B\!-\!I)_{\rm red}$ & 
 $(V\!-\!I)_{\rm blue}$ & $(V\!-\!I)_{\rm red}$ & 
 $f_{\rm red}$\tablenotemark{d} & $p$ \\
Limit & ($''$) & & (mag) & (mag) & (mag) & (mag) & (mag) &  & value \\ [0.5ex]
\tableline
\multicolumn{3}{c}{~} \\ [-2ex]     
$B<25.5$  &  50  &  92  & 0.10  & 1.61 & 1.91   &       &        & 0.70 & 0.010 \\ 
          & 100  & 135  & 0.10  & 1.60 & 1.91   &       &        & 0.64 & 0.000 \\
$B<26.5$  &  50  & 153  & 0.12  & 1.63 & 1.94   &       &        & 0.71 & 0.007 \\
          & 100  & 241  & 0.12  & 1.61 & 1.93   &       &        & 0.64 & 0.000 \\ 
$V<25.0$  &  50  & 111  & 0.06  &      &        & 0.92  & 1.08   & 0.80 & 0.023 \\ 
          & 100  & 157  & 0.06  &      &        & 0.92  & 1.09   & 0.72 & 0.000 \\
$V<26.5$  &  50  & 210  & 0.08  &      &        & 0.93  & 1.09   & 0.80 & 0.043 \\
          & 100  & 369  & 0.07  &      &        & 0.93  & 1.09   & 0.69 & 0.001 \\ [0.5ex]
\tableline
\end{tabular}

\tablenotetext{a}{Outer galactocentric radius of GC sample under consideration.} 
\tablenotetext{b}{Number of GCs in sample.} 
\tablenotetext{c}{Standard deviation of Gaussian fit by KMM test.} 
\tablenotetext{d}{Fraction of red GCs in color distribution indicated by KMM test.} 
\end{center}

\end{table*}

\begin{figure}
\epsscale{1.05}
\plotone{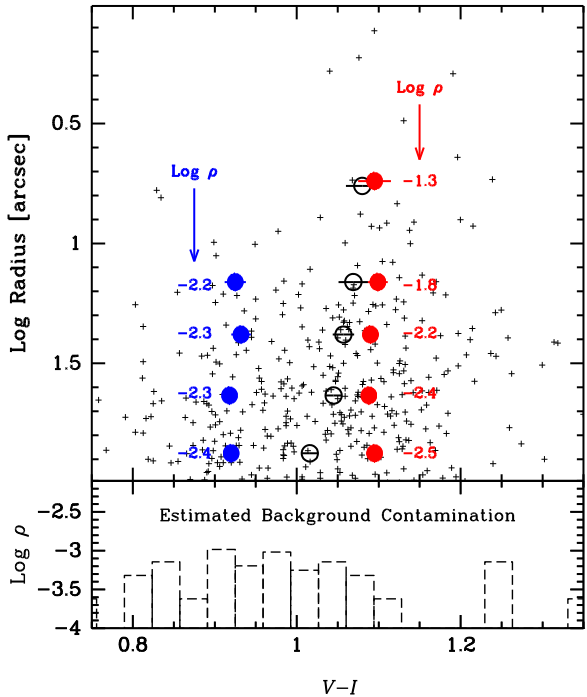}
\caption{
{\it Top panel}:\ \VI\ color versus logarithm of projected
galactocentric radius for GC 
candidates in NGC~3610 with $V \le 26.5$. The plus signs represent the
individual GC candidates. The large open black circles represent the mean colors
of GC candidates using bins of 0.25 in Log\,(radius) (except for the innermost
bin, which represents Log\,(radius) $\le$ 1). The large filled red and
blue circles represent the mean colors of the red and blue GC subpopulations,
respectively, using the same radial binning as for the black circles. Error
bars drawn for the large circles represent mean errors of the
mean. Values mentioned beside each red and blue circle indicate the
local surface number density of red and blue GC candidates (in arcsec$^{-2}$), 
respectively.  
{\it Bottom panel}: \VI\ color distribution of
`contaminating' compact background galaxies as estimated from the outer
regions of the image, expressed in surface number density units
(arcsec$^{-2}$). See text in Section~\ref{s:colors} for further details.  
\label{f:color_lograd}}
\end{figure}

The dependence of the color distribution of GC candidates on galactocentric
radius is illustrated in Figure~\ref{f:color_lograd}, which shows 
\VI\ color vs.\ projected galactocentric radius for all GC candidates with $V
\le 26.5$. In addition, Figure~\ref{f:color_lograd} shows average colors of
GC candidates in bins of log\,(radius) for three samples: {\it (i)\/} all GC
candidates, {\it (ii)\/} GC candidates in the `blue' subpopulation ($0.75 \le
V\!-\!I \le 1.00$), and {\it (iii)\/} GC candidates in the `red' subpopulation ($1.00 \le
V\!-\!I \le 1.30$). The average color of the full sample of GC candidates
gets bluer with increasing projected radius, as found before for many
`normal' ellipticals \citep[e.g.,][and references
therein]{ashzep98}. Fig.~\ref{f:color_lograd} also shows that
the average colors of the blue and red subpopulations stay constant with
increasing radius (see Fig.~\ref{f:color_lograd}), and that these results are
not affected by contamination by compact background galaxies. Hence, the 
overall radial color gradient is due to a more centrally concentrated radial
distribution of the  
red GCs with respect to their blue counterparts. This effect (which is quantified for  
NGC~3610 in Section~\ref{s:numdens} below) was predicted by the
`formation by dissipative merger' scenario of \citet{ashzep92}, and confirmed by
recent studies of giant ellipticals hosting populous GC systems with bimodal
color distributions \citep[e.g.,][]{geis+96,goud+01b,sikk+06}. 

\begin{figure*}
\epsscale{0.95}
\plotone{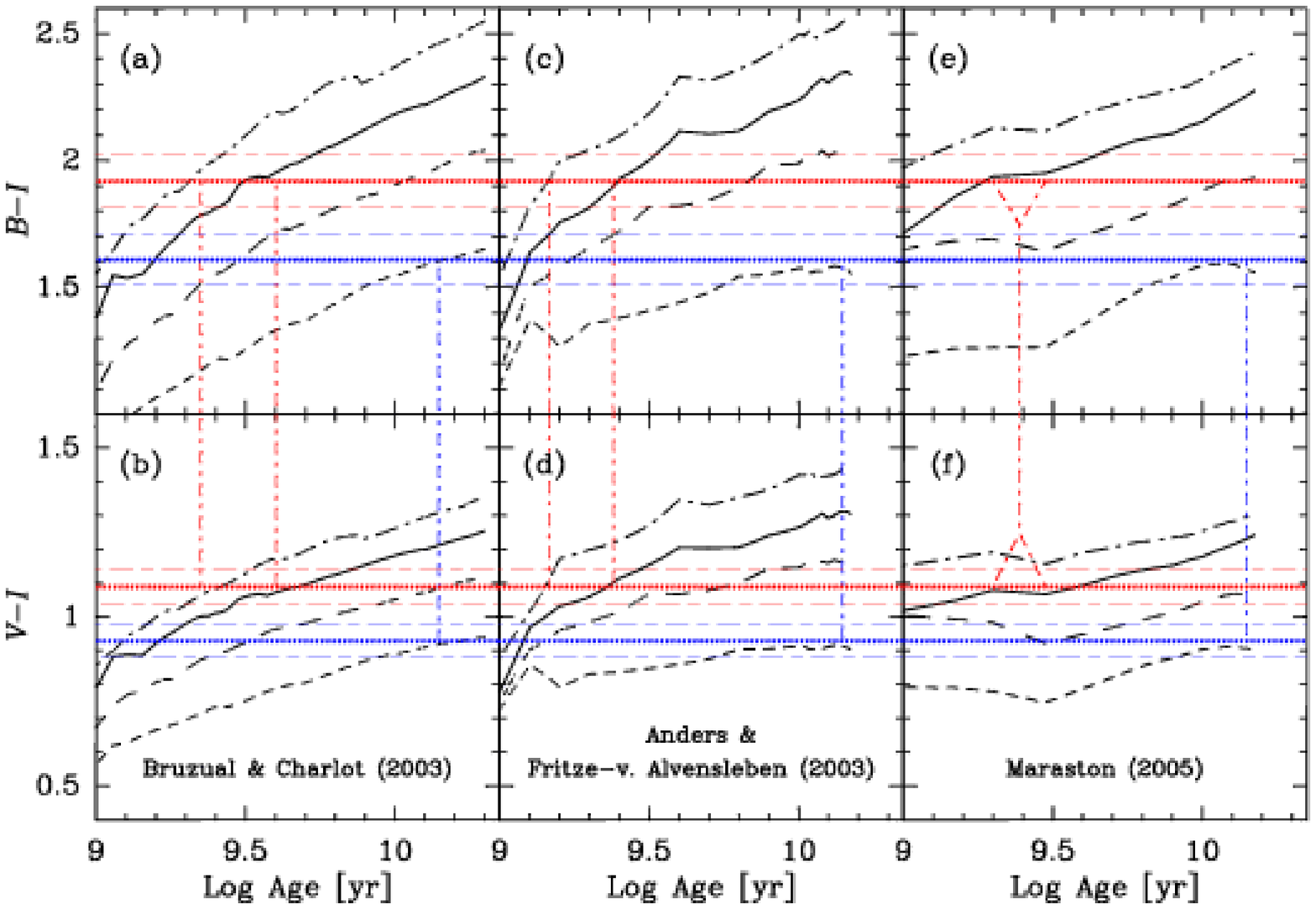}
\caption{
Time evolution of $B\!-\!I$ (top panels) and \VI\ (bottom panels) color indices 
 according to single-burst stellar population models for ages older than 1
 Gyr.  
Panels (a) and (b) show \citet{bc03} models, panels (c) and (d) show
 \citet{andfri03} models, and panels (e) and (f) show \citet{mara05}
 models. Model curves are plotted in black  for 
 the following metallicities:\ 0.02 solar
 {\it (short-dashed lines)}, 0.2 solar {\it (long-dashed lines)}, 1.0
 solar {\it (solid lines)}, and 2.5 solar {\it (dot-long-dashed lines)}.  
The horizontal red and blue dotted lines indicate the peak colors as
determined by the KMM test for the red and blue GC subpopulations,
respectively (see Sect.~\ref{s:colors}). The horizontal long-short-dashed
lines above and below the red and blue dotted lines indicate the 
$\pm 1 \sigma$ widths of the two Gaussians fitted to the bimodal color
distribution.  
The vertical blue dot-short-dashed lines in each panel indicate the 14 Gyr 
model age. The vertical red dot-short-dashed lines indicate the best-fit ages to
the red peak color for solar metallicity (right line) and for 2.5 solar
metallicity (left line), separately for each SSP model. The exception
 is the \citet{mara05} model where a {\it range\/} of ages (2\,--\,3 Gyr) fits
 the colors well for solar metallicity, and 2.5 solar metallicity does
 not fit the data well for ages $>$ 1 Gyr. Hence only one red vertical
 line is drawn for that case. 
\label{f:sspmodels}}
\end{figure*}

The peak colors of the blue and red subpopulations mentioned above are
compared with popular SSP models in Figure~\ref{f:sspmodels}. We considered
the models of \citet[][using the Salpeter initial mass function
(IMF)]{andfri03}, \citet[][using the Chabrier IMF]{bc03}, and \citet[][using
the Kroupa IMF]{mara05} for this purpose. As to the observed peak 
\BI\ and \VI\ colors of the blue GCs, Figure~\ref{f:sspmodels} shows that all
three models fit these quite well at an age of 14 Gyr and [$Z$/H] = $-$1.7, which
is consistent with the mean [$Z$/H] of old Galactic halo GCs
\citep[e.g.,][]{harr96}. Given that the luminosity function of the blue
GC subsystem is also consistent with that of the Galactic halo GCs (see
below), we consider the blue GC subsystem to consist of old, metal-poor GCs
for all intents and purposes. As to the red GC subsystem,
Figure~\ref{f:sspmodels} shows that its peak \BI\ and \VI\ colors are 
consistent with various age/metallicity combinations. These are discussed
in turn below.  

One option to explain the colors of the red GCs is that they are `old'
(age $\ga$ 10 Gyr) GCs with $-1.0 \la [Z/{\rm H}] \la -0.5$ (the exact
best-fit value for [$Z$/H] depending on the preferred SSP model), i.e., GCs
with properties similar to those of bulge GCs in our Galaxy
\citep[e.g.,][]{puzi+02,mara+03}, which may have been introduced into
NGC~3610 if (at least) one of the merger progenitor spiral galaxies
hosted a significant bulge component. 
The existence of old GCs among the `red' GC subsystem in NGC~3610 was shown
by \citet{stra+04}, who determined ages and metallicities of 11 GCs in
NGC~3610 from spectra obtained with the Keck I telescope, using
spectroscopic Lick indices. Based on SSP model fits to the Lick indices,
Strader et al.\ argued that 7 out of the 9 `red' GCs in their sample
were best fit by old ages ($>$ 10 Gyr) and metallicities similar to
those of GCs in the Galactic bulge. This option will be considered
further in interpreting the LF of red GCs in Section~\ref{s:LFs} below. 

\begin{figure}
\epsscale{1.075}
\plotone{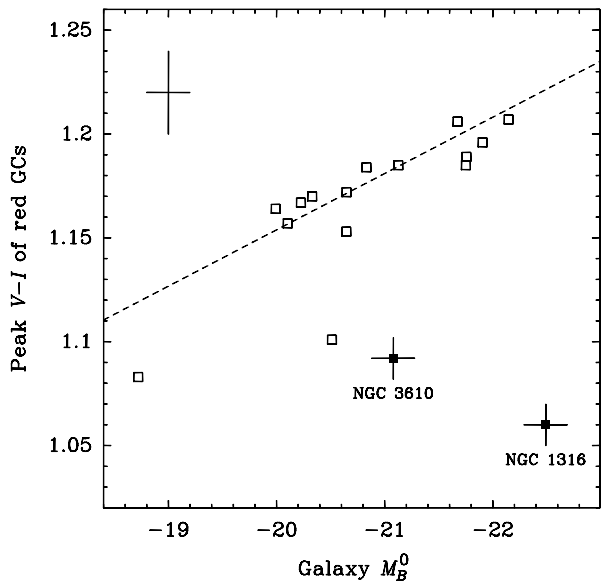}
\caption{
 Peak \VI\ color of the red GC subpopulation versus absolute $B$-band
 magnitude of parent galaxy among nearby elliptical galaxies. The open
 squares represent data from \citet{lars+01}, for which a typical error bar is shown at
 the top left of the figure. The dashed line represents a
 linear least-squares fit to the open squares. The data for NGC~3610 and NGC~1316
 \citep[data for the latter was taken from][]{goud+04} are 
 shown by filled squares. See discussion in Sect.~\ref{s:colors}. 
\label{f:Mb_vs_VmIred}}
\end{figure}

Another option that is consistent with the colors of the red GCs is that they
are of intermediate age ($\sim$\,1\,--\,4 Gyr) and metal rich ([$Z$/H] $\ga$
0.0). Support for this option is provided by several empirical results. First
of all, the LF of the red GC population in NGC~3610 is known to be well-fit
by a power law down to well beyond the turnover magnitude of known
`old' GC systems in giant ellipticals (W02; see also
Sect.~\ref{s:redLF} below).  
Second, the peak color of the red GCs in NGC~3610 is significantly {\it bluer\/} than
that of red GCs in `normal' elliptical galaxies at the luminosity of
NGC~3610. This is illustrated in Figure~\ref{f:Mb_vs_VmIred} which shows the relation 
between the peak \VI\ colors of the red GC subpopulation in nearby elliptical galaxies 
with clear bimodal color distributions versus the galaxies' absolute $B$-band magnitude 
from the study of \citet{lars+01}. 
Figure~\ref{f:Mb_vs_VmIred} clearly shows that these two properties correlate 
strongly with each other, which has been generally 
interpreted as a galaxy luminosity-metallicity relation similar to that among elliptical
galaxies themselves, suggesting a close connection between the physical
processes responsible for the formation of the red GCs and that of their
parent galaxies \citep[e.g.,][and references therein]{brostr06}. Assuming 
the same physical relation holds for NGC~3610 \citep[and for the
3-Gyr-old merger remnant NGC~1316:][]{goud+01b,goud+04}, the fact that  
the peak color of their red GCs are clearly offset to the blue from this
relation (see Figure~\ref{f:Mb_vs_VmIred}) must be due to a younger age
rather than a lower metallicity. Finally, the Lick indices of the
2 innermost red GCs in the spectroscopic sample of \citet{stra+04} were 
best fit by an age of $\sim$\,1.6 Gyr and metallicity [$Z$/H] $\sim$\,+0.3,
consistent with Lick-system analysis of spectra of the diffuse light of the
galaxy out to beyond its effective radius
\citep{howe+04}. Figure~\ref{f:sspmodels} shows that this age/metallicity   
combination fits the observed \BI\ and \VI\ colors quite well. 

\subsection{Cluster Luminosity Functions} \label{s:LFs}

\begin{figure*}
\epsscale{0.8}
\plotone{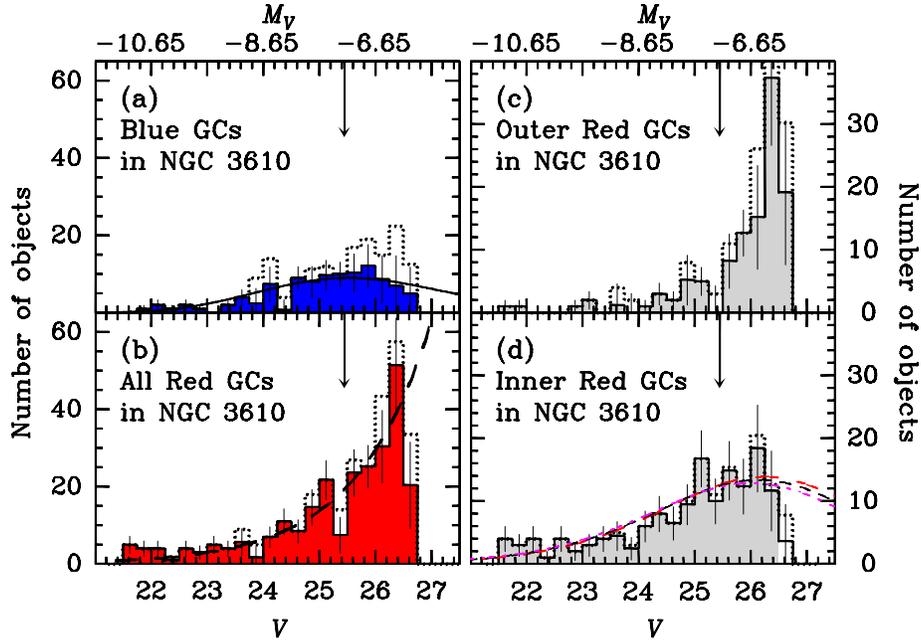}
\caption{
  $V$-band luminosity functions (LFs) of GC candidates in NGC~3610. 
  {\it Panel (a)}: LF of the full blue subsystem. 
  {\it Panel (b)}: LF of the full red subsystem. 
  {\it Panel (c)}: LF of the outer 50\%\ of the red subsystem. 
  {\it Panel (d)}: LF of the inner 50\%\ of the red subsystem. 
  Dotted histograms mark LFs corrected for incompleteness, and solid
  histograms mark LFs 
  corrected for incompleteness and compact background galaxies. Histograms are filled
  for magnitude bins brighter than the overall 50\%\ completeness limit of
  the sample plotted, and open beyond it. The smooth dashed curve in panel
  (b) is a power-law fit to the LF. Vertical arrows in all panels indicate the 
  typical turnover magnitude for `old' GC systems (i.e., $M_V = -7.2$). 
  Note the appearance of a flattening consistent with a turnover in the LF of
  the inner 50\% of the red subsystem, which is not present in the outer 50\%. 
  The smooth curves in panels (a) and (d) represent predicted LFs based on a
  combination of the \citet{falzha01} GC disruption models and the
  SSP models of \citet{mara05}. Curves are drawn for the
  following populations:\ ages of 1.5 (long-dashed line), 3.0 (short-dashed
  line), 6.0 Gyr (short-dash-dotted line), and 12 Gyr (solid line in panel
  (a)). \ZH\ = $-1.7$  is assumed for the 12 Gyr curve, whereas
  \ZH\ = 0.0 is assumed for the three younger ages. The
  normalization of all model curves in panels (a) and (d) was done by a
  least-squares fit to the underlying histograms.  
\label{f:blueredLFs}}    
\end{figure*}

The LFs of the blue and red GC subpopulations are shown in
Figure~\ref{f:blueredLFs}, and were determined as follows. We first
separated the GC candidates into `blue' ($0.75 \le V\!-\!I \le 0.98$) and
`red' ($1.03 \le V\!-\!I \le 1.30$) subsamples. Note that in this case, we
avoided the overlapping region ($0.98 < V\!-\!I < 1.03$) to minimize the
mixing of the two subsamples.  
A completeness correction was then applied to each GC by dividing 
its count (of 1) by the completeness fraction corresponding to its brightness and
background level. The latter fraction was calculated from the functions shown
in Figure~\ref{f:completeness}, using bilinear interpolation in log(background)
space.  
The LFs were further corrected for contamination by foreground stars and
compact background objects using the smoothed and scaled LF of the objects 
detected at galactocentric distances $R > 100''$ with the same color and
size criteria as the GC candidates at $R \le 100''$. The LFs of those outer
objects were smoothed by a five-bin median filter to avoid noise due to
small-number statistics and scaled to the area of the $R \le 100''$ region 
prior to subtraction. 
The observed LFs corrected for completeness are shown as dotted lines in
Figure~\ref{f:blueredLFs}, whereas the solid lines show the LFs after 
correction for contamination by compact background objects. 

As already reported by W02, the LFs of the (full) blue and red GC
subsystems in NGC~3610 differ significantly from each other. We 
discuss the LFs of the two subpopulations separately below. 

\subsubsection{The Blue Subpopulation} \label{s:blueLF}

The shape of the LF of the blue GC subsystem is similar to that typically
reported for GCs in `normal' giant ellipticals, i.e., approximately
Gaussian with a turnover magnitude $M_V \sim -7.2$ (e.g., the
compilation of Whitmore [\citeyear{whit97}] found $M_V = -7.21 \pm 0.26$). A
Gaussian fit 
to the LF of the blue GC subsystem of NGC 3610 results in a peak of $M_V
= -7.21 \pm 0.15$ and a standard deviation $\sigma = 1.22 \pm
0.18$. While the latter is consistent with that found among `normal'
giant early-type galaxies (e.g., Kundu \& Whitmore [2001] found values between
1.1\,--\,1.5 mag), it is significantly larger than that found for NGC~3610
by W02, who found $\sigma = 0.66 \pm 0.19$. We ascribe this difference to a 
larger field of view and a larger sample of GC
candidates in the current study relative to W02.  

\subsubsection{The Red Subpopulation} \label{s:redLF}

The LF for the full red GC system is consistent with a power law down to the
50\% completeness limit of the data. A weighted least-squares fit of
$\phi(L) dL \propto L^{\alpha} dL$ to the LF yields an exponent $\alpha = -1.78 \pm
0.06$, consistent with the value found by W02 ($\alpha = -1.78 \pm 0.05$). While this
confirms the nature of the bulk of the red GC population of NGC~3610 as having 
formed during a recent star-forming event, it does not provide significant
evidence to suggest an evolution of the LF from a power law to a Gaussian.   

However, disruption timescales of the main dynamical effects thought to be
responsible for the Gaussian shape of the GC LFs of `normal', old 
galaxies through preferential depletion of low-mass GCs 
\citep{vesp01,falzha01,baumak03} depend on galactocentric distance. The
stronger tidal field in the central regions yields stronger tidal shocks
than in the outskirts, while also imposing a more stringent zero-energy
surface on the GCs so that stars can be more easily removed from the GCs
through two-body relaxation \citep[see also][]{gneost97,gned97}.   
Hence, the appearance of a turnover in the GC LF can be expected to first
occur in the inner parts of the GC system. This was first shown to be the
case by \citet{goud+04} for the 3-Gyr-old merger remnant NGC~1316, where
the inner half of the red GC system showed a significant flattening of
the GC LF beyond the turnover expected (and found)
for the blue, metal-poor GC system. In contrast, the outer half of the
red GC system still rose strongly all the way to the detection limit. 
With the 580 GC candidates in our ACS images of NGC~3610, we are now in a
position to test this finding for this intermediate-age merger remnant with
high enough statistical significance as well. To do so, we divide the red
GC system into two equal-size parts sorted by projected galactocentric
radius. For our sample, this division ends up at 46\farcs7, or 4.8 kpc from
the galaxy center. The resulting LFs are shown in panels (c) and (d) of 
Figure~\ref{f:blueredLFs}. The LF of the {\it outer\/} half of the red GC system
still rises all the way to the detection limit, whereas {\it the LF of the
  inner half of the red GC system shows a clear flattening consistent with a
  turnover at $\sim$\,1.0 mag fainter than that of the blue GC
  system}. These properties are quite similar to those found for the red GC
system in NGC~1316 \citep{goud+04}. 

\begin{figure*}
\epsscale{0.8}
\plotone{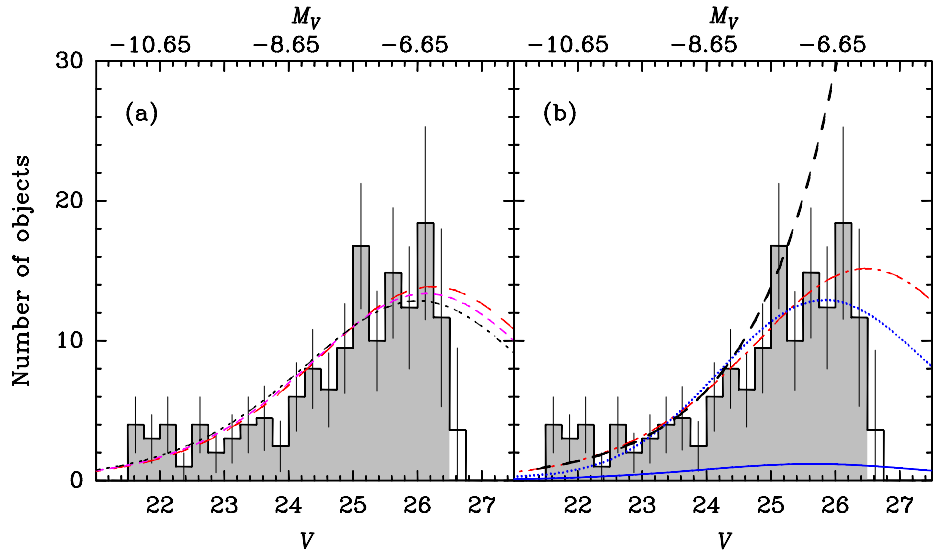}
\caption{
  $V$-band LFs of the red GC candidates in NGC~3610. Both panels show the
  same LF as in Figure~\ref{f:blueredLFs}d. 
  {\it Panel (a)}: 
  FZ01 model curves are drawn for the following populations of solar
  metallicity (using the \citet{mara05} SSP models):\ ages of 1.5
  (long-dashed line), 3.0 (short-dashed line), and 6.0 Gyr
  (short-dash-dotted line).  
  {\it Panel (b)}: FZ01 model curves are drawn for the following
  populations: Age = 1.5 Gyr and \ZH\ = +0.3 (long-dash-dotted line)
  and Age = 12 Gyr and \ZH\ = $-$0.7 (dotted line). The normalization
  of all aforementioned model curves was done by a least-squares fit
  to the underlying histogram. To illustrate the estimated
  contribution to the red GC LF by a population of GCs associated with
  the bulge of a giant spiral galaxy similar to the Milky Way, the solid
  line represents a FZ01 model 
  curve for Age = 12 Gyr and \ZH\ = $-$0.7 for a population of 50 red
  GCs (see discussion in Sect.~\ref{s:redLF}). For reference, a
  power-law fit to the full red GC population (cf.\
  Figure~\ref{f:blueredLFs}b) is shown as a long-dashed curve in panel (b). 
\label{f:redLFs}}    
\end{figure*}

We compare these results with predictions of GC disruption models by
\citet[][hereafter FZ01]{falzha01} in Figure~\ref{f:blueredLFs}. The FZ01 
models, whose initial GC velocity distribution involves a radial anisotropy distribution 
that is consistent with kinematics of halo stars in the solar neighborhood,  
showed that an initial power law or Schechter mass function evolves in 12 Gyr
to a peaked mass distribution similar to that observed for the Milky Way GC
system. To predict the shapes of {\it luminosity\/} functions as a function
of age, we use the FZ01 model that uses a power-law initial mass function  
in combination with the SSP models of \citet{mara05}. We
assume \ZH\ = $-$1.7 for the 12 Gyr old population (similar to the
median metallicity of halo GCs in our Galaxy), while we use \ZH\ =
0.0 for populations of ages 1.5, 3, and 6 Gyr.  
Prior to plotting, the amplitude of all resulting model LFs were normalized
to the observed data down to the 50\% completeness limit by weighted
least-square minimization. The individual RMS values of these fits to the 
histograms are listed in Table~\ref{t:LFfits}. 
The peak magnitudes of the 1.5, 3, and 6 Gyr model LFs of solar metallicity
span a range of only 0.3 mag in $V$, due to the counteracting effects of
disruption of (preferentially low-mass) GCs and luminosity fading of the
surviving GCs (see also W02). Hence the LF shapes of these three models are
quite similar to one another. However, as discussed in Sect.~\ref{s:colors},
recent spectroscopic investigations indicated that the intermediate-age 
GC population is of age $\sim$\,1.5 Gyr and has [$Z$/H] $\sim +0.3$, as
found for two bright, inner red GCs as well as the integrated galaxy light
\citep{stra+04,howe+04}. This possibility is illustrated in
Figure~\ref{f:redLFs}, which compares the models shown in 
Figure~\ref{f:blueredLFs}d with the combination of the 1.5 Gyr FZ01 model
and the \citet{mara05} model of 1.5 Gyr and [$Z$/H] = +0.3. Formally, the
least-squares fit of this model to the LF indeed yields the lowest RMS
residuals of all models for intermediate ages considered here.  
%
For reference, Figure~\ref{f:redLFs} also shows a fit of a 12 Gyr FZ01 model
with a metallicity similar to that of an average Galactic bulge GC
\citep[\ZH\ = $-0.7$, see][]{puzi+02,mara+03} to the LF of the inner 50\% of
the red GCs. As Table~\ref{t:LFfits} shows, this model does not fit the data
nearly as well as any of the intermediate-age models. Furthermore, if the majority of the
red GCs would be $\sim$\,12 Gyr old and of moderate metallicity, one
would expect this model to fit the red GC LF independent of
galactocentric radius, as it does for the metal-rich GCs associated
with the Galactic bulge as well as for red GCs in ``normal'' giant
ellipticals. In contrast, the LF of the 50\% outermost red GCs is
consistent with a power law as shown in Figure~\ref{f:blueredLFs}c. 

\begin{table}
\begin{center}
\caption{Fits of FZ01 mass functions to luminosity function of the
  inner 50\% of the red GCs\tablenotemark{a}. \label{t:LFfits}}   
\small
\begin{tabular*}{4cm}{@{\extracolsep{\fill}}crc@{}}
\multicolumn{3}{c}{~} \\ [-2.5ex]   
\tableline \tableline
\multicolumn{3}{c}{~} \\ [-1.8ex]                                                
Age\tablenotemark{b} & [$Z$/H]\tablenotemark{c} & rms \\
 (Gyr) & (dex) & error \\ [0.5ex]
\tableline
\multicolumn{3}{c}{~} \\ [-2ex]     
1.5  & +0.3\, & 2.38 \\
1.5  &  0.0\, & 2.43 \\
3.0  &  0.0\, & 2.48 \\
6.0  &  0.0\, & 2.57 \\
\llap{1}2.0 & $-$0.7\, & 3.12 \\ [0.5ex]
\tableline
\end{tabular*}

\tablenotetext{a}{GCs with $V \le 26.5$ have been considered in the fits.} 
\tablenotetext{b}{Assumed age of red GCs.} 
\tablenotetext{c}{Assumed [$Z$/H] of red GCs.} 
\end{center}

\end{table}

\begin{figure}
\epsscale{1.15}
\plotone{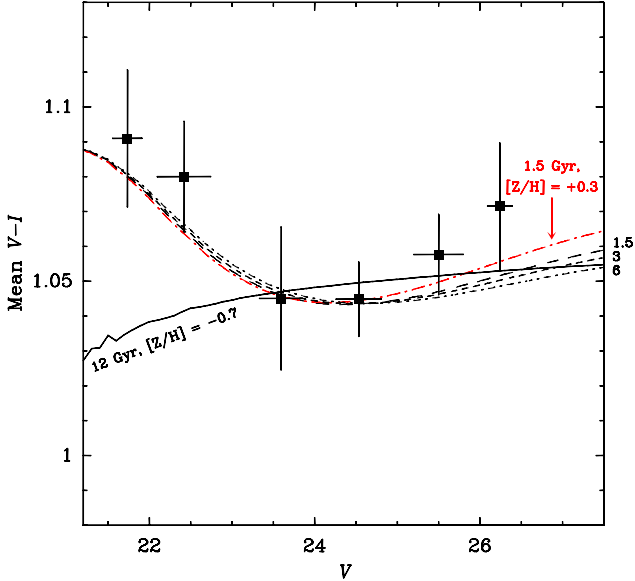}
\caption{
  Mean \VI\ colors of GC candidates as a function of $V$ magnitude in the
  inner 46\farcs7. The squares represent measured mean \VI\ colors after
  statistical subtraction of compact background galaxies (see discussion in
  Sect.~\ref{s:redLF}). The dashed lines indicate mean \VI\ colors as
  predicted by the FZ01 models normalized to the LFs shown in
  Figures~\ref{f:blueredLFs} and \ref{f:redLFs}. The line types are the same
  as in those figures, and ages (in Gyr) are indicated on the right-hand
  side of the diagram. Note that the fit of the intermediate-age models to
  the data is significantly better than that of the old model with
  metallicity similar to that of bulge GCs in our Galaxy (\ZH\ = $-0.7$). 
\label{f:meancolors}}    
\end{figure}

\begin{figure}
\epsscale{0.9}
\plotone{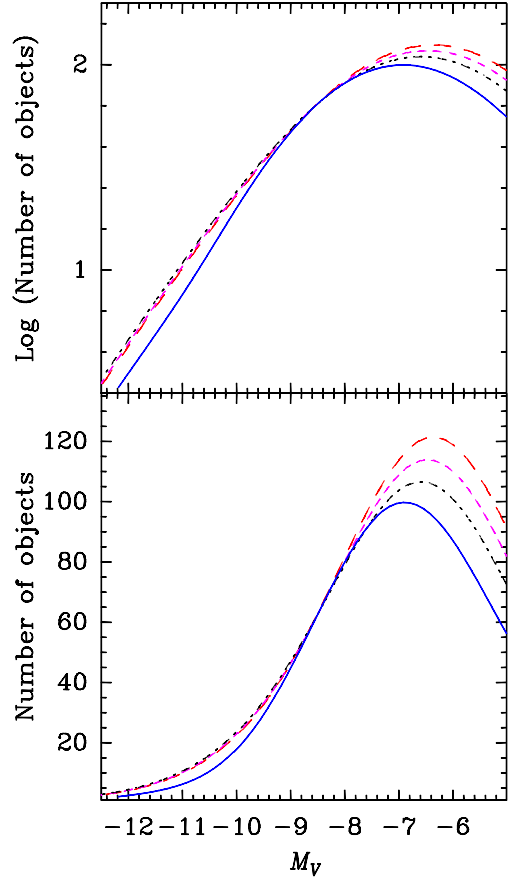}
\caption{ 
Predicted $V$-band GC LFs in relative logarithmic (top panel) and linear 
(bottom panel) units. Then 
drawn model curves are \citet{falzha01} GC 
disruption models for the following age/metallicity combinations (using the 
\citet{mara05} SSP models): Age = 1.5 Gyr and \ZH\ = 0.0 (red long-dashed
line), Age = 3 Gyr and \ZH\ = 0.0 (magenta short-dashed line), Age = 6 Gyr and
\ZH\ = 0.0 (black short-dash-dotted line) and Age = 12 Gyr and \ZH\ = $-$0.7
(blue solid line). These age/metallicity combinations fit the color of the red
peak in the GC color distribution fairly well and are also plotted in
Fig.~\ref{f:redLFs}. The curves are normalized to one another at $M_V = -8.5$,
in the steep part of the high-luminosity wing of the LF. Note that the
luminosity fading predicted by the SSP models of Maraston is somewhat faster
than the GC disruption rate predicted by the Fall \& Zhang models at high
masses.\label{f:gclfs_comparing}}   
\end{figure}

Another way to test whether the red GC subpopulation is more likely to be
dominated by an intermediate-age population with [$Z$/H] $\ga 0$ or by an
old bulge-like population with \ZH\,$\sim$\,$-0.7$ is by comparing the observed
mean colors of GCs as a function of their luminosity with the predictions of
the cluster disruption models fit to the LFs of the blue and red GCs (i.e.,
the curves in Figures~\ref{f:blueredLFs}a and \ref{f:redLFs}). 
This comparison is shown in Figure~\ref{f:meancolors} for the inner
half of the GC population ($r \le$ 46\farcs7), for which we showed that
the FZ01 models fit the LFs of both the blue and the red GC subpopulations 
quite well. To calculate mean colors as predicted by the models, we
assume \VI\ = 0.93 and \VI\ = 1.09 for the blue and red
subpopulations, respectively (cf.\ Table~\ref{t:kmmresults}). The
observed mean \VI\ colors are calculated for GC candidates with $0.7
\le V\!-\!I \le 1.5$. Figure~\ref{f:meancolors} shows that the
observed behavior of the mean GC color as function of $V$ magnitude
(the red clusters dominating at both the bright and faint ends of the
LF) is clearly better fit by metal-rich intermediate-age models than
by the old model with bulge-GC-like metallicity (\ZH\ =
$-0.7$). (Incidentally, the 1.5 Gyr model with [$Z$/H] $\sim +0.3$ yields
the best fit to the observed mean colors, just as it does to the LF of
the red GCs.) This result is mainly due to the lack of high-luminosity red GCs
in the 12 Gyr FZ01 model with \ZH\ = $-$0.7 relative to the situation for the
models of younger age. The underlying reason is that at high GC masses and old
ages, age fading from one luminosity bin to the next (lower) one is somewhat faster
than the disruption rate of GCs in that lower luminosity (or mass) bin. This is
demonstrated in Figure~\ref{f:gclfs_comparing} in which we plot the same model
LFs, normalized to each other at $M_V = -8.5$ (in the steep part of the
high-luminosity wing of the LF), both on logarithmic and linear scales. The
bright end of the LF moves to lower luminosities between 6 and 12 Gyr.  

From the above considerations, it seems most likely that the red
GC subpopulation in NGC~3610 is dominated by an intermediate-age
population with \ZH$\;\ga\;0.0$ rather than an old population with
\ZH$\;\sim -0.7$. As Figure~\ref{f:redLFs} shows, the location
of the turnover found for the inner half of the red GC system is in
rather good agreement with those of the FZ01 models for
intermediate ages and solar metallicity.   
It thus seems reasonable to suggest that the LF of this second-generation GC
system will evolve further to become consistent with the LFs of red GCs in
`normal' giant ellipticals.  

\begin{figure}
\centerline{\includegraphics[width=8.4cm]{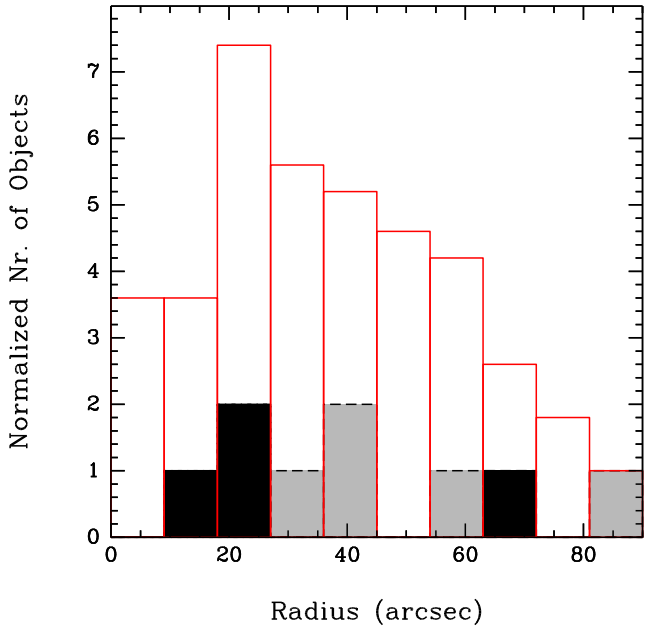}}
\caption{
Distribution of projected galactocentric radii of red GCs studied by 
\citet[][dashed lines, hatched histogram]{stra+04} compared with that of the 
red GC candidates in the current study (solid lines, open histogram). The latter 
histogram was normalized to the former in the outermost radial bin. The dark-hatched 
histogram (in black) highlights the 4 GCs for which \citet{stra+04} obtained spectra 
with S/N ratios $>$ 10 per \AA. 
\label{f:radiushist}} 
\end{figure}

Our results seem to differ from those of \citet{stra+04} who argued 
(based on spectra taken using the Keck I telescope) that 7 out of
their 9 red GCs in NGC~3610 
are old ($\ga 10$ Gyr) rather than of intermediate age. To address this apparent 
discrepancy, we offer the following insights. {\it (i)\/} Most red GCs
targeted by \citet{stra+04}  
were located in the outer regions of NGC~3610, whereas the majority of the bright 
red GCs is centrally located, as might be expected for GCs formed in a relatively recent 
starburst event associated with a dissipative galaxy merger. This is 
illustrated in Figure~\ref{f:radiushist} which compares a histogram of
the projected radii of the red GCs targeted by \citet{stra+04} with
that of the red GC system from the current study. 
{\it (ii)\/} 
The faintness of the sources and the relatively high galaxy
background caused many of Strader et al.'s GC spectra to be of marginal
quality. Applying a commonly assumed absolute minimum requirement on
the signal-to-noise (S/N) ratio to allow a useful age/metallicity analysis
\citep*[S/N $>$ 10 per \AA, see
e.g.][]{kunt+02,lars+03,beas+04,puzi+02,puzi+05,puzi+06}, Strader et al.'s
results amount to four red GCs with spectra of 
adequate S/N of which two are old and two are of intermediate age. 
(This excludes clusters W22 and W28 [i.e., \#\,20 and \#\,26 in the current study, 
respectively], which are clearly blue GCs as judged from the ACS data, cf.\ 
Table~\ref{t:phot}.) While we believe a small sample of 4 GCs with
adequate spectral data does not allow one to draw strong conclusions
regarding the fraction of red GCs being old vs.\ of intermediate age,
it is clear from the \citet{stra+04} study that old GCs do exist among
the red GC subsample in NGC~3610. As these old, red GCs presumably
stem from the progenitor galaxies that merged to form NGC~3610, it
seems likely that at least one of those galaxies possessed a
significant bulge. For example, the Milky Way (a giant Sbc galaxy)
hosts about 20\,--\,30 old, metal-rich GCs associated with its bulge
\citep[e.g.,][]{minn+95}. The LF expected for a population of 50 such
GCs according to the FZ01 model for an age of 12 Gyr is illustrated in
Figure~\ref{f:redLFs}b as a solid curve. Note that the
contribution of such a population is negligible relative to the
observed numbers of red GCs, especially since
Figure~\ref{f:redLFs}c only refers to the inner 50\% of the red GC
subsystem.   

\subsection{Radial Distributions and Specific Frequencies}
\label{s:numdens}  

The specific frequency of GCs, 
\[ S_N \equiv N_{\rm GC} \, 10^{0.4\,(M_V + 15)} \]
\citep{harvdb81}, i.e., the number of star clusters per galaxy luminosity in
units of $M_V$ = $-$15, is an important parameter of GC systems in 
galaxies. $S_N$ is known to increase along the Hubble sequence,
with mean values ranging from $\left< S_N \right> = 0.5$ for Sc spirals
\citep{goud+03,chan+04} to $2.9$ for `normal' giant
ellipticals outside galaxy clusters, albeit with considerable
galaxy-to-galaxy scatter \citep[e.g.,][]{ashzep98,brostr06}.  

The intermediate-age merger remnant NGC~3610 is a well-suited probe to
test whether the difference in specific frequency between spirals and
(giant) ellipticals can be accounted for by secondary populations of clusters 
created during (gas-rich) mergers. We know that its body has already
settled down to a typical $r^{1/4}$ surface brightness law, we have a
relatively good estimate of the age of the major merger during which
the secondary population of GCs were created \citep[both from the
properties of the GCs themselves and from deep spectra of the
diffuse galaxy light;][]{howe+04}, and it occurred long enough ago for
dynamical evolution of the system to have already had a significant
impact. Hence, estimates of evolutionary effects ought to be less
uncertain than for younger mergers. In the remainder of this section, we
derive the current value of $S_N$ for NGC~3610 and estimate its future
evolution.  

\subsubsection{Radial Number Density Distribution of Blue and Red
  Clusters} 
The radial distribution of the surface number density of blue and red GCs in
NGC~3610 was derived from the ACS data as follows. To avoid
incompleteness-related issues, we only considered GCs brighter than
the 50\% completeness limit in the central regions ($V < 24.5$) for
this exercise. We calculated the completeness-corrected number of blue
and red GCs in four annuli, logarithmically spaced in projected
galactocentric radius. Next, we divided the number of clusters in each
annulus by the appropriate area coverage to yield a surface density (taking
into account the limited azimuthal sky coverage of the data at large
radii). Finally, we subtracted the surface number density of foreground stars and
compact background galaxies for both blue and red GCs, measured as detailed
in  Sect.~\ref{s:colors} above.   

\begin{figure}
\includegraphics[angle=-90,width=8.3cm]{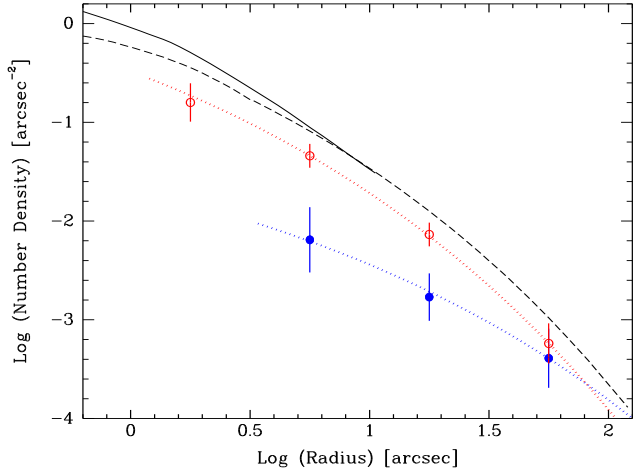}
\caption{
Radial number density profiles for the GC system of
 NGC~3610 down to $V$ = 24.5. Data have been corrected for
 completeness, areal coverage, and contamination by compact background
 galaxies as explained in the text. Filled circles represent the blue
 GCs while open circles represent the red GCs. Dotted lines drawn through
 the two sets of circles represent de Vaucouleurs' 
 $r^{1/4}$ profile fits to the respective GC number density data. The dashed line
 represents the surface brightness profile of the integrated
 $V$-band light of the galaxy \citep[taken from][]{goud+94a} shifted
 by an arbitrary constant, while the solid line does the same for the
 integrated galaxy light as measured from the ACS $V$-band
 images. \label{f:rhoplot}} 
\end{figure}

Figure~\ref{f:rhoplot} shows the derived number density distributions 
of the blue and red GCs, along with the surface brightness profile of
the galaxy light for comparison purposes. The surface number density profile 
of the red GCs is largely consistent with the galaxy light profile 
(the innermost surface number density value is slightly below the latter),
whereas that of the blue GCs is significantly flatter than the galaxy
light profile. Quantitatively, least-square de Vaucouleurs'
(\citeyear{dev48}) $r^{1/4}$ 
profile fits to the GC number density profiles yield $r_e = 128 \pm
24$ arcsec (or $20.5 \pm 3.8$ kpc) for the blue GCs and $r_e = 19.6
\pm 0.2$ arcsec (or $3.14 \pm 0.03$ kpc) for the red GCs. While
radial number density profiles of red GCs in giant ellipticals are
typically found to be steeper than those of blue GCs
\citep[e.g.,][]{ashzep98}, the difference in slope is typically
smaller than that found here for NGC~3610. In particular, the number
density profile of the red GCs is significantly closer to that of the
galaxy light in NGC~3610 than it is in `normal' giant elliptical
galaxies \citep{forb+98,forb+06,dirs+03,dirs+05,bass+06}, where the red GCs
are more depleted towards the center than the galaxy light. This
difference is likely due to the cumulative effects of two-body
relaxation and tidal shocking of GCs (both of which are most effective in the
central regions as discussed in Sect.\ \ref{s:redLF} above) and dynamical
friction  \citep[the latter only for the most massive GCs,][]{vesp01}. The
more moderate flattening of the number density of the
red GC system of NGC~3610 towards the center relative to that of `normal' 
giant ellipticals is consistent with the intermediate-age nature of NGC~3610.   

\subsubsection{Specific Frequencies of Blue and Red Clusters} 

The number of GCs in the formula for the specific frequency $S_N$ was
originally defined for `old' GC systems which show LFs consistent with
a Gaussian, namely as twice the number of GCs brighter than the GCLF
turn-over magnitude \citep{harvdb81}. This is the procedure we follow
for the {\it blue\/} GCs, for which the turnover magnitude $M_V = -7.2$
corresponds to a mass of $2.4 \times 10^5$ M$_{\odot}$ according to the \citet{mara05}
SSP model for an age of 14 Gyr and $[Z/\mbox{H}] = -1.7$. As the {\it red\/}
GCs in NGC~3610 are most likely predominantly of intermediate age rather than
old, we use an alternative definition of $S_N$ for the red GCs: We calculate
twice the number of red GCs {\it more massive\/} than the mass associated
with the turnover magnitude of the Galactic GC system, taking into account
the uncertainties of the assumed age and metallicity for the red GCs. 
To this end, we use the SSP tables of \citet{mara05} and age/metallicity
combinations ranging from Age = 1.5 Gyr and $[Z/\mbox{H}] = +0.3$ to Age = 4
Gyr and $[Z/\mbox{H}] = 0.0$ (cf.\ Figure~\ref{f:sspmodels}). 


Using the $r^{1/4}$ profile fits to the GC number density profiles mentioned
above, we calculate the total number of blue and red GCs more massive than
$2.4 \times 10^5$ M$_{\odot}$. For the purpose of $S_N$ calculations, we separate
blue and red GCs at \VI\ = 1.00 and we assume a distance uncertainty
of 15\%. Results are listed in Table~\ref{t:S_N}.  

\begin{table*}
\begin{center}
\caption{Results of Specific Frequency calculations. \label{t:S_N}}  
\footnotesize
\begin{tabular*}{1.0\textwidth}{@{\extracolsep{\fill}}crccccccccc@{}}
\multicolumn{3}{c}{~} \\ [-2.5ex]   
\tableline \tableline
\multicolumn{3}{c}{~} \\ [-1.8ex]
Age\tablenotemark{a} & [$Z$/H]\tablenotemark{b} & 
 N$_{\rm blue}$\tablenotemark{c} & N$_{\rm red}$\tablenotemark{d} & 
 $S_{N, {\rm blue, now}}$\tablenotemark{e} & $S_{N, {\rm red, now}}$\tablenotemark{f} & 
 $S_{N, {\rm tot, now}}$\tablenotemark{g} & N$_{\rm red, old}$\tablenotemark{h} & 
 $S_{N, {\rm blue, old}}$\tablenotemark{i} & $S_{N, {\rm red, old}}$\tablenotemark{j} & 
 $S_{N, {\rm tot, old}}$\tablenotemark{k} \\ [0.5ex]
\tableline
\multicolumn{8}{c}{~} \\ [-2ex]     
1.5 & +0.3 & $376 \pm 136$ & $412 \pm 40$ & $0.71 \pm 0.36$ & $0.78 \pm 0.28$ & $1.49 \pm 0.59$ & $371 \pm 36$ & $2.09 \pm 1.21$ & $2.06 \pm 0.77$ & $4.15 \pm 1.76$ \\
2.0 &  0.0 & $376 \pm 136$ & $396 \pm 40$ & $0.71 \pm 0.36$ & $0.75 \pm 0.27$ & $1.46 \pm 0.58$ & $356 \pm 36$ & $2.09 \pm 1.21$ & $2.02 \pm 0.75$ & $4.11 \pm 1.75$ \\
3.0 &  0.0 & $376 \pm 136$ & $304 \pm 36$ & $0.71 \pm 0.36$ & $0.57 \pm 0.21$ & $1.28 \pm 0.52$ & $274 \pm 32$ & $2.09 \pm 1.21$ & $1.52 \pm 0.58$ & $3.61 \pm 1.61$ \\
4.0 &  0.0 & $376 \pm 136$ & $268 \pm 32$ & $0.71 \pm 0.36$ & $0.51 \pm 0.19$ & $1.22 \pm 0.51$ & $241 \pm 29$ & $2.09 \pm 1.21$ & $1.34 \pm 0.51$ & $3.43 \pm 1.56$ \\ [0.5ex]
\tableline
\end{tabular*}
\tablenotetext{a}{Assumed age of red GCs in Gyr.} 
\tablenotetext{b}{Assumed [$Z$/H] of red GCs in dex.} 
\tablenotetext{c}{Total number of blue GCs around NGC~3610.} 
\tablenotetext{d}{Total number of red GCs around NGC~3610.} 
\tablenotetext{e}{Present-day specific frequency of blue GCs.} 
\tablenotetext{f}{Present-day specific frequency of red GCs.} 
\tablenotetext{g}{Total present-day specific frequency of GCs.} 
\tablenotetext{h}{Number of red GCs at age of 10 Gyr, corrected for disruption.} 
\tablenotetext{i}{Specific frequency of blue GCs at age of 10 Gyr.} 
\tablenotetext{j}{Specific frequency of red GCs at age of 10 Gyr.} 
\tablenotetext{k}{Total specific frequency of GCs at age of 10 Gyr.} 
\end{center}

\end{table*}

\begin{figure}
\centerline{\includegraphics[width=8cm]{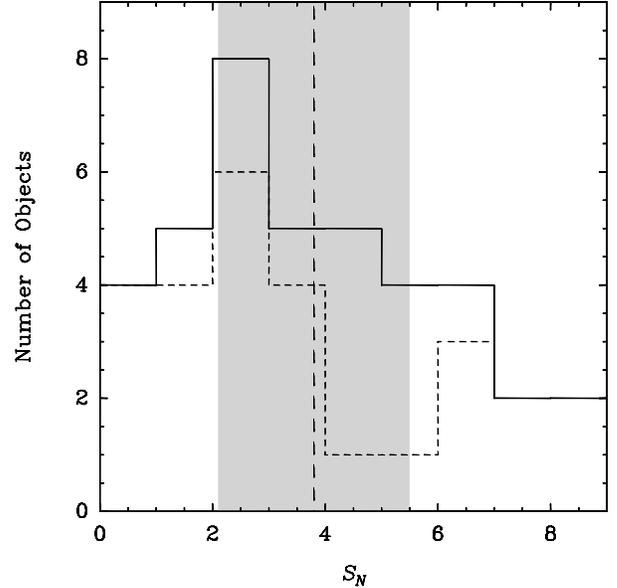}}
\caption{
 Histogram of specific frequencies ($S_N$ values) of GCs in elliptical
 galaxies in the compilation of \citet{ashzep98}. Central dominant galaxies
 in galaxy clusters have been excluded. The solid lines represent all
 ellipticals in the compilation, whereas the short-dashed lines represent all
 ellipticals outside galaxy clusters. For comparison, the specific frequency
 of GCs in NGC~3610 ($S_N = 3.8 \pm 1.7$) is indicated by a vertical long-dashed
 line; a shaded region around it indicates its uncertainty. \label{f:SnHisto}}   
\end{figure}

Using a total $V$-band magnitude of 10.84 for NGC~3610 \citep{RC3} and hence 
$M_V^0 = -21.81$ at the assumed distance of NGC~3610, the 
total {\it present-day\/} $S_N = 1.36 \pm 0.58$. To allow a proper comparison
of this estimate with those of `normal', old ellipticals, this value still
needs to be corrected for two evolutionary effects: 
\begin{enumerate}
\item 
Luminosity fading of the diffuse light of NGC~3610. Treating  
the diffuse light of NGC~3610 as a SSP, we assume an age of 3
$\pm$ 1 Gyr for NGC~3610 and an age of 10 Gyr for a `normal' elliptical. This
luminosity fading amounts to 1.17 $\pm$ 0.31 mag in the $V$ band. 
An alternative option is that the diffuse light of NGC~3610 is produced by a
combination of an old and a young population. Using the \citet{mara05} models at
solar metallicity, we find that the color of the diffuse light (\BI\ =
1.92 and \VI\ = 1.11, see Goudfrooij et al.\ \citeyear{goud+94a}) can be modeled as 
a combination of a 10-Gyr-old component (73\% by luminosity) plus a 1-Gyr-old 
component (27\% by
luminosity). 
In this case, the luminosity fading of the young component to an age of 10 Gyr would 
amount to 0.80 mag, which is near the low end of the range of fading values 
calculated for the SSP age of 3 $\pm$ 1 Gyr mentioned above.   
\item 
The fraction of red GCs more massive than $2.4 \times 10^5$ M$_{\odot}$ that will
be disrupted during this period. To estimate this fraction, we 
use the FZ01 model mass functions that involve a power-law initial mass
function and find that the number of GCs more massive than $2.4 \times
10^5$ M$_{\odot}$ decreases by 12.0\% between ages of 1.5 and 12 Gyr,
and by 9.8\% between ages of 3 and 12 Gyr. For the current estimate,
we adopt a value of 10\%.  
\end{enumerate}
Taking the corrections mentioned above into account, our final estimate of
the total specific frequency of GCs in NGC~3610 at an age of 10 Gyr is $S_N
\simeq 3.8 \pm 1.7$. The fraction of red GCs contributing to this value is
$0.45 \pm 0.05$. While these values should be checked using a future deep
wide-field imaging study of NGC~3610 (e.g., the study of NGC~7814 by Rhode \& Zepf 
[\citeyear{rhozep03}] showed that one can overestimate $S_N$ by $\sim$\,20--30\% 
when extrapolating results from \HST\ images to large radii), we do note that this value 
of $S_N$ is consistent with typical values found for giant ellipticals in loose
groups\footnote{NGC~3610 is the brightest galaxy of a 
loose group of galaxies \citep[LGG\,234, 5 members;][]{garc93}.}. This is 
illustrated by Figure~\ref{f:SnHisto} which compares the value for NGC~3610 found
here with a histogram of $S_N$ values for `normal' ellipticals 
\citep[taken from the compilation of][which does not include
NGC~3610]{ashzep98}. Quantitatively, the mean and standard deviation of specific 
frequencies of ellipticals outside galaxy clusters in the latter compilation is 
$S_N = 2.9 \pm 1.9$. 

\section{Summary and Conclusions}   \label{s:conc}

We have presented a study of globular cluster candidates in NGC~3610, a
giant elliptical galaxy that is very likely an intermediate-age merger remnant, using
deep {\it BVI\/} images taken with the ACS camera on board the {\it Hubble Space
  Telescope}. Our main results are as follows. 
\begin{enumerate}
\item
The globular clusters show a clear bimodal color distribution, both in \BI\
and in \VI. The peak colors of the blue subpopulation are \BI\ = 1.61 $\pm$ 0.01and \VI\ =
0.93 $\pm$ 0.01, and those of the red subpopulation are \BI\ = 1.92 $\pm$ 0.01 and \VI\ = 
1.09 $\pm$ 0.01. The
colors of the blue clusters are consistent with those of blue clusters in other elliptical
galaxies as well as of halo clusters in our Galaxy, while the colors of the red clusters
are significantly bluer than the colors indicated by the relation of red cluster
color with galaxy luminosity among `normal' elliptical galaxies. Given the
additional fact that the brightest red clusters are up to 0.5 mag brighter
than the brightest blue clusters in NGC~3610, we interpret this as evidence
for a {\it younger age\/} of the red clusters in NGC~3610 relative to normal
giant ellipticals rather than a significantly lower metallicity.  
\item
The luminosity function of the blue, metal-poor subpopulation of clusters in
NGC~3610 is similar to a Gaussian with $M_V^0 \simeq -7.2$ and $\sigma \simeq
1.2$, consistent with values typical for old blue cluster populations in
normal elliptical galaxies. 
\item 
The luminosity function of the full red subpopulation
of clusters in NGC~3610 is consistent with a power law $\phi(L) dL \propto
L^{\alpha} dL$ with $\alpha = -1.78 \pm 0.06$, which is consistent with the
result of W02 using earlier WFPC2 data. However, the luminosity function of
the {\it inner half\/} of the red cluster subsystem shows a clear
flattening consistent with a turnover at $M_V^0 \simeq -6.2$. A comparison with 
predictions of
GC disruption models in conjunction with simple stellar population models at
various ages and metallicities shows that both the luminosity function of the
inner half of the red cluster subsystem and the relation of mean cluster
color with luminosity are best fit by a population of metal-rich clusters of
intermediate age ($\sim 1.5-3$ Gyr).  
As similar results were found for the red GCs in the 3-Gyr-old merger remnant
NGC~1316 \citep{goud+04}, this result supports the
assertion that dynamical evolution of metal-rich GC populations formed in
gas-rich galaxy mergers indeed changes their LF properties to become
consistent with those of the red, metal-rich GC subsystems that are
ubiquitous in `normal' giant ellipticals.  
\item
The surface number density profile of the blue GC subpopulation is much
more radially extended than the surface brightness profile of the
galaxy light, similar to the situation in  `normal', old elliptical
galaxies. In contrast, the radial profile of the red GC subpopulation
follows that  of the galaxy light almost all the way into the galaxy
center. The difference in steepness between the radial profile of the
red GC population and that of the galaxy light is smaller than
reported for `normal', old elliptical galaxies. The combination of
this result with the finding that the red GCs and the galaxy light
appear to have ages and metallicities consistent with each other
suggests that most of the red GCs were formed during the same event as the
bulk of the field stars in NGC~3610 and that they have suffered
little disruption after violent relaxation with respect to the
situation in `normal' ellipticals. 
\item
We determine the specific frequency of clusters in NGC~3610 and
find a present-day value of $S_N = 1.4 \pm 0.6$. Using published age
estimates for the diffuse light of NGC~3610 as well as cluster disruption
models, we estimate that this value will increase to $S_N = 3.8 \pm 1.7$
at an age of 10 Gyr, which is consistent with typical $S_N$ values of
`normal' elliptical galaxies in loose groups. 
\end{enumerate}

Our results support a picture in which the formation process of giant
ellipticals with significant populations of metal-rich GCs was similar to that
of gas-rich galaxy mergers observed today.

\paragraph*{Acknowledgments.}~We acknowledge useful interactions with 
Enrico Vesperini, Rupali Chandar, and Thomas Puzia. We thank the
anonymous referee for insightful comments which improved the paper. Support for
{\it HST\/} Program GO-9409 was provided by NASA through a grant from
the Space Telescope Science Institute, which is operated by the Association of
Universities for Research in Astronomy, Inc., under NASA contract NAS5--26555.


\begin{thebibliography}{}
\bibitem[Anders \& Fritze-v.\ Alvensleben(2003)]{andfri03}
Anders, P., \& Fritze-v.\ Alvensleben, U., 2003, \aap, 401, 1063
\bibitem[Ashman \& Zepf(1992)]{ashzep92}
Ashman, K. M., \& Zepf, S. E., 1992, \apj, 384, 50
\bibitem[Ashman \& Zepf(1998)]{ashzep98}
Ashman, K. M., \& Zepf, S. E., 1998, ``Globular Cluster Systems''
 (Cambridge: Cambridge University Press) 
\bibitem[Ashman et al.(1994)Ashman, Bird, \& Zepf]{ashm+94}
Ashman, K. M., Bird, C. M., \& Zepf, S. E., 1994, \aj, 108, 2348
\bibitem[Barnes(2002)]{barn02}
Barnes, J. E., 2002, \mnras, 333, 481
\bibitem[Bassino et al.(2006)Bassino, Richtler, \& Dirsch]{bass+06}
Bassino, L. P., Richtler, T., \& Dirsch, B., 2006, \mnras, 367, 156
\bibitem[Bastian et al.(2006)]{bast+06}
Bastian, N., Saglia, R. P., Goudfrooij, P., Kissler-Patig, M., Maraston, C., 
 Schweizer, F., \& Zoccali, M., 2006, \aap, 448, 881
\bibitem[Baumgardt(1998)]{baum98}
Baumgardt, H., 1998, \aap, 330, 480 
\bibitem[Baumgardt \& Makino(2003)]{baumak03}
Baumgardt, H., \& Makino, J., 2003, \mnras, 340, 227
\bibitem[Beasley et al.(2004)]{beas+04}
Beasley, M. A., Brodie, J. P., Strader, J., Forbes, D. A., Proctor,
 R. N., Barmby, P., \& Huchra, J. P., 2004, \aj, 128, 1623
\bibitem[Brodie \& Strader(2006)]{brostr06} 
Brodie, J. P.,  \& Strader, J., 2006, \araa, 44, 193
\bibitem[Bruzual \& Charlot(1993)]{bc93} 
Bruzual, G. A., \& Charlot, S., 1993, \apj, 405, 538 
\bibitem[Bruzual \& Charlot(2003)]{bc03} 
Bruzual, G. A., \& Charlot, S., 2003, \mnras, 344, 1000
\bibitem[Burstein \& Heiles(1984)]{burhei84} 
Burstein, D., \& Heiles, C.\ 1984, \apjs, 54, 33
\bibitem[Burstein et al.(1987)]{burs+87} 
Burstein, D., Davies, R. L., Dressler, A., Faber, S. M., Stone, R. P. S.,
 Lynden-Bell, D., Terlevich, R., \& Wegner, G., 1987, \apjs, 64, 601
\bibitem[Chandar et al.(2004)Chandar, Whitmore, \& Lee]{chan+04}
Chandar, R., Whitmore, B. C., \& Lee, M. G., 2004, \apj, 611, 220
\bibitem[Cohen et al.(2003)Cohen, Blakeslee, \& C\^ot\'e]{cohe+03}
Cohen, J. G., Blakeslee, J. P., \& C\^ot\'e, P., 2003, ApJ, 592, 866
\bibitem[de Vaucouleurs(1948)]{dev48}
de Vaucouleurs, G., 1948, Ann.\ d'Astrophys., 11, 247
\bibitem[de Vaucouleurs et al.(1991)]{RC3}
de Vaucouleurs, G., de Vaucouleurs, A., Corbin, H. G., Buta, R. J.,
 Paturel, G., \& Fouque, P., 1991, ``Third Reference Catalogue of Bright
 Galaxies'' (New York: Springer) 
\bibitem[Dirsch et al.(2003)]{dirs+03}
Dirsch, B., Richtler, T., Geisler, D., Forte, J. C., Bassino, L. P.,
 \& Gieren, W. P., 2003, \aj, 125, 1908
\bibitem[Dirsch et al.(2005)Dirsch, Schuberth, \& Richtler]{dirs+05}
Dirsch, B., Schuberth, Y., \& Richtler, T., 2005, \aap, 433, 43
\bibitem[Fall \& Zhang(2001)]{falzha01}
Fall, S. M., \& Zhang, Q., 2001, ApJ, 561, 751 (FZ01)
\bibitem[Forbes et al.(1997)Forbes, Brodie, \& Grillmair]{forb+97}
Forbes, D. A., Brodie, J. P., \& Grillmair, C. J., 1997, \aj, 113, 1652
\bibitem[Forbes et al.(1998)]{forb+98}
Forbes, D. A., Grillmair, C. J., Williger, G. M.., Elson, R. A. W., \&
 Brodie, J. P., 1998, \mnras, 293, 325
\bibitem[Forbes et al.(2001)]{forb+01}
Forbes, D. A., Beasley, M. A., Brodie, J. P., \& Kissler-Patig, M., 2001,
 ApJ, 563, L143
\bibitem[Forbes et al.(2006)]{forb+06}
Forbes, D. A., S\'anchez-Bl\'azquez, P., Phan, A. T. T., Brodie,
 J. P., Strader, J., \&  Spitler, L., 2006, \mnras, 366, 1230
\bibitem[Garcia(1993)]{garc93}
Garcia, A. M., 1993, \aaps, 100, 47
\bibitem[Geisler et al.(1996)Geisler, Lee, \& Kim]{geis+96}
Geisler, D., Lee, M. G., \& Kim, E., 1996, \aj, 111, 1529
\bibitem[Gnedin(1997)]{gned97}
Gnedin, O. Y., 1997, \apj, 487, 663
\bibitem[Gnedin \& Ostriker(1997)]{gneost97}
Gnedin, O. Y., \& Ostriker, J. P., 1997, \apj, 474, 223
\bibitem[Goudfrooij et al.(1994a)]{goud+94a}
Goudfrooij, P., N{\o}rgaard-Nielsen, H. U., Hansen, L., 
   J{\o}rgensen, H. E., de Jong, T., \& van den Hoek, L. B., 1994a,
   \aaps, 104, 179
\bibitem[Goudfrooij et al.(1994b)]{goud+94b}
Goudfrooij, P., N{\o}rgaard-Nielsen, H. U., Hansen, L., \& 
   J{\o}rgensen, H. E., 1994b, \aaps, 105, 341 
\bibitem[Goudfrooij et al.(2001a)]{goud+01a}
Goudfrooij, P., Mack, J., Kissler-Patig, M., Meylan, G., \& 
 Minniti, D., 2001a,  \mnras, 322, 643
\bibitem[Goudfrooij et al.(2001b)]{goud+01b}
Goudfrooij, P., Alonso, M. V., Maraston, C., \& Minniti, D., 2001b,
 \mnras, 328, 237 
\bibitem[Goudfrooij et al.(2003)]{goud+03}
Goudfrooij, P., Strader, J. M., Brenneman, L., Kissler-Patig, M., \& Huizinga, 
  E. J., 2003, \mnras, 343, 665
\bibitem[Goudfrooij et al.(2004)]{goud+04}
Goudfrooij, P., Gilmore, D., Schweizer, F., \& Whitmore, B. C., 2004,
 \apjl, 613, L121 
\bibitem[Harris(1996)]{harr96}
Harris, W. E., 1996, \aj, 112, 1487 
\bibitem[Harris \& van den Bergh(1981)]{harvdb81}
Harris, W. E., \& van den Bergh, S., 1981, \aj, 86, 1627
\bibitem[Harris et al.(2006)]{harr+06}
Harris, W. E., Whitmore, B. C., Karakla, D., Oko\'n, W., Baum, W. A.,
 Hanes, D. A., \& Kavelaars, J. J., 2006, \apj, 636, 90
\bibitem[Holtzman et al.(1992)]{holt+92}
Holtzman, J. A., Faber, S. M., Shaya, E. J., et al.\ 1992, \aj, 103, 691
\bibitem[Howell et al.(2004)]{howe+04}
Howell, J. H., Brodie, J. P., Strader, J., Forbes, D. A., \& Proctor,
 R., 2004, AJ, 128, 2749
\bibitem[Idiart et al.(2002)Idiart, Michard, \& de Freitas Pacheco]{idia+02}
Idiart, T. P., Michard, R., \& de Freitas Pacheco, J. A., 2002, \aap, 383, 30
\bibitem[Knapp et al.(1989)]{knap+89}
Knapp, G. R., Guhathakurta, P., Kim, D. W., \& Jura, M. A., 1989, \apjs, 70, 329
\bibitem[Kundu \& Whitmore(2001)]{kunwhi01}
Kundu, A., \& Whitmore, B. C., 2001, AJ, 121, 1888
\bibitem[Kuntschner et al.(2002)]{kunt+02}
Kuntschner, H., Ziegler, B. L., Sharples, R. M., Woorthey, G., \&
 Fricke, K. J., 2002, \aap, 395, 761
\bibitem[Larsen et al.(2001)]{lars+01}
Larsen, S. S., Brodie, J. P., Huchra, J. P., Forbes, D. A., \& Grillmair,
  C. J., 2001, \aj, 121, 2974
\bibitem[Larsen et al.(2003)]{lars+03}
Larsen, S. S., Brodie, J. P., Beasley, M. A., Forbes, D. A., 
 Kissler-Patig, M., Kuntschner, H., \& Puzia, T. H., 2003, \apj, 589, L81
\bibitem[Maraston(2005)]{mara05}
Maraston, C., 2005, \mnras, 362, 799
\bibitem[Maraston et al.(2001)]{mara+01}
Maraston, C., Kissler-Patig, M., Brodie, J. P., Barmby, P., \& Huchra, J. P.,
 2001, \aap, 370, 176
\bibitem[Maraston et al.(2003)]{mara+03}
Maraston, C., Greggio, L., Renzini, A., Ortolani, S., Saglia, R. P.,
 Puzia, T. H., \& Kissler-Patig, M., 2003, \aap, 400, 823
\bibitem[Maraston et al.(2004)]{mara+04}
Maraston, C., Bastian, N., Saglia, R. P., Kissler-Patig, M.,
 Schweizer, F., \& Goudfrooij, P., 2004, \aap, 416, 467
\bibitem[McLachlan \& Basford(1988)]{mclbas88}
McLachlan, G. J., \& Basford, K. E., 1998, Mixture Models: Inference
 and Application to Clustering (New York: M. Dekker) 
\bibitem[Michard(2005)]{mich05}
Michard, R., 2005, \aap, 429, 819
\bibitem[Minniti(1995)]{minn+95}
Minniti, D., 1995, AJ, 109, 1663
\bibitem[Peng et al.(2006)]{peng+06}
Peng, E. W., Jord\'an, A., C\^ot\'e, P., et al., 
  2006, \apj, 639, 95
\bibitem[Prieto \& Gnedin(2006)]{prigne06}
Prieto, J. L., \& Gnedin, O. Y., 2006, submitted to ApJ (astro-ph/0608069)
\bibitem[Puzia et al.(2002)]{puzi+02}
Puzia, T. H., Saglia, R. P., Kissler-Patig, M., Maraston, C., Greggio,
 L., Renzini, A., \& Ortolani, S., 2002, \aap, 395, 45
\bibitem[Puzia et al.(2005)]{puzi+05}
Puzia, T. H., Kissler-Patig, M., Thomas, D., Maraston, C., Saglia, R. P.,
 Bender, R., Goudfrooij, P., \& Hempel, M., 2005, \aap, 439, 997
\bibitem[Puzia et al.(2006)Puzia, Kissler-Patig, \& Goudfrooij]{puzi+06}
Puzia, T. H., Kissler-Patig, M., \& Goudfrooij, P., 2006, \apj, 648, 383
\bibitem[Rhode \& Zepf(2003)]{rhozep03}
Rhode, K. L., \& Zepf, S. E., 2003, \aj, 126, 2307
\bibitem[Riess \& Mack(2004)]{riemac04}
Riess, A., \& Mack, J., 2004, Instrument Science Report ACS 2004-06
 (Baltimore: STScI) 
\bibitem[Salpeter(1955)]{salp55} 
Salpeter, E. E., 1955, \apj, 121, 161
\bibitem[Schlegel et al.(1998)Schlegel, Finkbeiner, \& Davis]{schl+98}
Schlegel, D. J., Finkbeiner, D. P., \& Davis, M.\ 1998, \apj, 500, 525
\bibitem[Schweizer(1987)]{schw87}
Schweizer, F., 1987, in: ``Nearly Normal Galaxies'', ed.\ S.\ M.\
 Faber (Springer: New York), 18
\bibitem[Schweizer(2002)]{schw02}
Schweizer, F., 2002, in: ``Extragalactic Star Clusters'', eds.\ D. Geisler,
 E. K. Grebel, \& D. Minniti (ASP: San Francisco), 630 
\bibitem[Schweizer \& Seitzer(1992)]{schsei92}
Schweizer, F., \& Seitzer, P., 1992, AJ, 104, 1039
\bibitem[Schweizer \& Seitzer(1998)]{schsei98}
Schweizer, F., \& Seitzer, P., 1998, AJ, 116, 2206
\bibitem[Seitzer \& Schweizer(1990)]{seisch90}
Seitzer, P., \& Schweizer, F., 1990, in: ``Dynamics and Interactions of
 Galaxies'', ed.\ S.\ M.\ Faber (Springer: New York), 18
\bibitem[Scorza \& Bender(1990)]{scoben90}
Scorza, C., \& Bender, R., 1990, \aap, 235, 49
\bibitem[Sikkema et al.(2006)]{sikk+06}
Sikkema, G., Peletier, R. F., Carter, D., Valentijn, E. A., \&
 Balcells, M., 2006, \aap, 458, 53
\bibitem[Sirianni et al.(2005)]{siri+05}
Sirianni, M., Jee, M. J., Ben\'{\i}tez, N., et al.\ 2005, \pasp, 117, 1049
\bibitem[Stetson(1987)]{stet87}
Stetson, P. B., 1987, \pasp, 99, 191 
\bibitem[Strader et al.(2004)Strader, Brodie, \& Forbes]{stra+04}
Strader, J., Brodie, J. P., \& Forbes, D. A., 2004, AJ, 127, 295
\bibitem[Strader et al.(2006)]{stra+06}
Strader, J., Brodie, J. P., Spitler, L., \& Beasley, M. A., 2006, AJ,
 132, 2333
\bibitem[Tonry et al.(2001)]{tonr+01}
Tonry, J. L., Dressler, A., Blakeslee, J. P., et al., 2001, \apj, 546,  681 
\bibitem[van den Bergh(1995)]{vdb95a}
van den Bergh, S., 1995, \apj, 450, 27
\bibitem[Vesperini(1998)]{vesp98}
Vesperini, E., 1998, \mnras, 299, 1019
\bibitem[Vesperini(2001)]{vesp01}
Vesperini, E., 2001, \mnras, 322, 247
\bibitem[Vesperini \& Zepf(2003)]{veszep03}
Vesperini, E., \& Zepf, S. E., 2003, \apj, 587, L97
\bibitem[Whitmore(1997)]{whit97}
Whitmore, B. C., 1997, in: ``The Extragalactic Distance Scale'', eds. M. Livio,
 M. Donahue, \& N. Panagia (Baltimore: STScI), 254
\bibitem[Whitmore et al.(1997)]{whit+97}
Whitmore, B. C., Miller, B. W., Schweizer, F., \& Fall, S. M., 1997,
\aj, 114, 1797
\bibitem[Whitmore et al.(1999)]{whit+99}
Whitmore, B. C., Zhang, Q., Leitherer, C., Fall, S. M., Schweizer, F.,
 \& Miller, B. W., 1999, AJ, 118, 1551
\bibitem[Whitmore et al.(2002)]{whit+02}
Whitmore, B. C., Schweizer, F., Kundu, A., \& Miller, B. W., 2002, AJ,
 124, 147
\bibitem[Zepf \& Ashman(1993)]{zepash93}
Zepf, S. E., Ashman, K. M., 1993, \mnras, 264, 611
\end{thebibliography}
\end{document}